\documentclass[prd, nofootinbib, notitlepage, superscriptaddress, preprintnumbers]{revtex4-1}

\usepackage{graphicx}
\usepackage{amsmath,bm}
\usepackage[T1]{fontenc}
\usepackage{rotating}
\usepackage{multirow}
\usepackage{hhline}
\usepackage{float}

\usepackage{color}
\definecolor{naviBlue}{RGB}{0,0,128}
\usepackage[colorlinks  = true,
            linkcolor   = naviBlue,
            urlcolor    = naviBlue,
            citecolor   = naviBlue,
            anchorcolor = naviBlue]{hyperref}

\newcommand{\gsim}{\ensuremath{\raisebox{-0.4 em}{ $\overset{>}{\sim}$ }}}
\newcommand{\lsim}{\ensuremath{\raisebox{-0.4 em}{ $\overset{<}{\sim}$ }}}

\newcommand{\Tproj}{T_i}
\newcommand{\pb}{{\bar{p}}}

\newcommand{\Tpbar}{T_{\bar{p}}}

\newcommand{\sS}{\sqrt{s}}
\newcommand{\xR}{x_\mathrm{R}}

\newcommand{\pT}{p_{\mathrm{T}}}

\newcommand{\sv}{\langle\sigma v \rangle}

\newcommand{\diff}{\mathrm{d}}

\begin{document}

\title{Scrutinizing the evidence for dark matter in cosmic-ray antiprotons}

\author{Alessandro Cuoco}
\email{cuoco@physik.rwth-aachen.de}
\affiliation{Universit\'e Grenoble Alpes, USMB, CNRS, LAPTh, F-74940 Annecy, France}
\affiliation{Institute for Theoretical Particle Physics and Cosmology, RWTH Aachen University, Sommerfeldstr.\ 16, 52056 Aachen, Germany}
\author{Jan Heisig}
\email{jan.heisig@uclouvain.be}
\affiliation{Centre for Cosmology, Particle Physics and Phenomenology (CP3), Universit\'e catholique de Louvain, Chemin du Cyclotron 2, B-1348 Louvain-la-Neuve, Belgium}
\author{Lukas Klamt}
\email{klamt@physik.rwth-aachen.de}
\affiliation{Institute for Theoretical Particle Physics and Cosmology, RWTH Aachen University, Sommerfeldstr.\ 16, 52056 Aachen, Germany}
\author{Michael Korsmeier}
\email{korsmeier@physik.rwth-aachen.de}
\affiliation{Institute for Theoretical Particle Physics and Cosmology, RWTH Aachen University, Sommerfeldstr.\ 16, 52056 Aachen, Germany}
\affiliation{Dipartimento di Fisica, Universit\`a di Torino, Via P. Giuria 1, 10125 Torino, Italy}
\affiliation{Istituto Nazionale di Fisica Nucleare, Sezione di Torino, Via P. Giuria 1, 10125 Torino, Italy}
\author{Michael Kr\"amer}
\email{mkraemer@physik.rwth-aachen.de}
\affiliation{Institute for Theoretical Particle Physics and Cosmology, RWTH Aachen University, Sommerfeldstr.\ 16, 52056 Aachen, Germany}

\preprint{LAPTH-052/18, TTK-19-09, CP3-19-08}

\begin{abstract}

Global fits of primary and secondary cosmic-ray (CR) fluxes measured by AMS-02 have great potential to study CR propagation models and search for exotic sources of antimatter such as annihilating dark matter (DM). Previous studies of AMS-02 antiprotons revealed a possible hint for a DM signal which, however, could be affected by systematic uncertainties. To test the robustness of such a  DM signal, in this work we systematically study two important sources of uncertainties: the antiproton production cross sections needed to calculate the source spectra of secondary antiprotons and the potential correlations in the experimental data, so far not provided by the AMS-02 Collaboration. To investigate the impact of cross-section uncertainties we perform global fits of CR spectra including a covariance matrix determined from nuclear cross-section measurements. As an alternative approach, we perform a joint fit to both the CR and cross-section data. The two methods agree and show that cross-section uncertainties have a small effect on the CR fits and on the significance of a potential DM signal, which we find to be at the level of $3\sigma$. Correlations in the data can have a much larger impact. To illustrate this effect, we determine possible benchmark models for the correlations in a data-driven method. The inclusion of correlations strongly improves the constraints on the propagation model and, furthermore, enhances the significance of the DM signal up to above $5\sigma$. Our analysis demonstrates the importance of providing the covariance of the experimental data, which is needed to fully exploit their potential.
\end{abstract}

\maketitle

%===================================================================
\section{Introduction}\label{sec::intro}
%===================================================================

Antimatter in cosmic rays (CRs), and in particular antiprotons, have been extensively investigated as a powerful means to search for exotic CR sources, such as dark matter (DM) annihilation in the Galaxy 
\cite{Bergstrom:1999jc,Donato:2003xg,Bringmann:2006im,Donato:2008jk,Fornengo:2013xda,Hooper:2014ysa,Pettorino:2014sua,Boudaud:2014qra,Cembranos:2014wza,Cirelli:2014lwa,Bringmann:2014lpa,Giesen:2015ufa,Evoli:2015vaa,Evoli:2011id,Reinert:2017aga,Cui:2018klo}.
The recent very accurate measurement of the CR antiproton flux by the \mbox{AMS-02} experiment~\cite{Aguilar:2016kjl} has significantly increased the sensitivity to a DM signal. A DM contribution as low as about $10\%$ of the antiproton flux can now in principle be detected, provided that 
the theoretical and experimental systematic uncertainties are under control at that level. Indeed, strong limits on heavy DM have been derived from global CR fits~\cite{Cuoco:2017iax}. At the same time, the data have also revealed a tentative signal of DM, corresponding to a DM mass of around 40--130 GeV and a thermal annihilation cross section, $\left\langle \sigma v \right\rangle \sim 3 \times 10^{-26}$~cm$^3$/s~\cite{Cuoco:2016eej,Cui:2016ppb,Cuoco:2017rxb}. This signal, if confirmed, is compatible with a DM interpretation of the Galactic center $\gamma$-ray excess (GCE) for a variety of annihilation channels. 
It is also expected to provide a detectable signal in antideuterons~\cite{Korsmeier:2017xzj}, and it is compatible
with a variety of different beyond-the-standard-model scenarios~\cite{Clark:2017fum,Cirelli:2016rnw,Cuoco:2017rxb,Cline:2017lvv,Jia:2017kjw,Arcadi:2017vis,Li:2017nac,Escudero:2017yia}.
However, given the  small experimental errors, several important sources of systematic uncertainties, which before could be neglected, now become increasingly important and need to be further investigated.

One such uncertainty concerns the predictions for the antiproton cross sections, needed to model antiproton production through scattering of CR protons and helium with the interstellar medium (ISM) in the Galactic disk. Recent progress in the determination of antiproton cross sections from nuclear experimental data~\cite{Winkler:2017xor}  has been found to have a significant impact on the DM interpretation of cosmic rays~\cite{Reinert:2017aga}. 
A second important source of uncertainty are possible correlations in the AMS-02 data, which are dominated by systematic uncertainties 
in most parts of the determined energy range.\footnote{The proton and helium fluxes are dominated by systematic uncertainties in the whole energy range from 1~GeV to 3~TeV, while in the antiproton-to-proton ratio systematic uncertainties are dominant from 1.8 to 50~GeV only.}
 The AMS-02 Collaboration has only released absolute systematic uncertainties, without providing information about their correlations. 
However, most of the systematic uncertainties are expected to exhibit sizable correlations. 

In this paper we systematically study the above-mentioned sources of uncertainties. We examine their impact on the significance of the potential DM excess in the CR antiproton data and hence scrutinize the robustness of this finding. Furthermore, we shed light on the impact of systematic uncertainties on the parameter determination of the CR propagation model.
To this end, we first reproduce the finding of~\cite{Cuoco:2016eej} in an updated setup, using the most recent cross-section parametrization
from~\cite{Korsmeier:2018gcy} and an improved treatment of solar modulation. This fit is considered as the default setup providing the reference for the following investigations.

We  study the impact of uncertainties in the antiproton production cross section following two approaches. The first approach is similar to the one taken in~\cite{Reinert:2017aga}. We incorporate the cross-section uncertainties in the CR fit by including a covariance matrix
extracted from a separate fit to nuclear measurements.
In the second approach we perform a joint fit of the CR propagation and antiproton cross-section parameters to the AMS-02 and the nuclear data. Such a joint fit provides important information about possible correlation between propagation and cross-section parameters.

We investigate the potential impact of correlations in the experimental data by assuming 
that the systematic uncertainties consist of three components: a part which is uncorrelated, a part
correlated over a certain number of neighboring rigidity bins, and a part which is fully correlated.
We determine these properties in a data-driven method, which allows us to constrain the viable range of the various components and to define four corresponding benchmark models. We perform global fits for these four benchmark models, each with and without a primary source of antiprotons from DM\@. 

The paper is structured as follows. In Sec.~\ref{sec:cr} we review CR propagation and highlight some new features of our setup which extend the standard treatment. Furthermore, we discuss solar modulation and justify our choice to omit low-rigidity data from the global fit. In Sec.~\ref{sec:setup} we detail the numerical implementation and describe our default setup for propagation and the corresponding results, considering the case with and without an additional source of antiprotons from DM annihilation. The uncertainties from antiproton cross sections are discussed in Sec.~\ref{sec::xs}, following the two approaches mentioned above. Finally, in Sec.~\ref{sec::corr} we discuss the potential impact of correlated errors in the AMS-02 data, following again two different methods. We conclude in Sec.~\ref{sec::conclusion}. 

%===================================================================
\section{Cosmic-ray propagation and solar modulation}\label{sec:cr}
%===================================================================

The propagation of CRs through the Galaxy can be described by the well-known diffusion equation for the phase-space densities of CRs~\cite{StrongMoskalenko_CR_rewview_2007}:
\begin{eqnarray}
  \label{eqn::propagationEquation}
  \frac{\partial \psi_i (\bm{x}, p, t)}{\partial t} = 
    q_i(\bm{x}, p) &+&  
    \bm{\nabla} \cdot \left(  D_{xx} \bm{\nabla} \psi_i - \bm{V} \psi_i \right) \nonumber \\ 
     &+&  \frac{\partial}{\partial p} p^2 D_{pp} \frac{\partial}{\partial p} \frac{1}{p^2} \psi_i - 
    \frac{\partial}{\partial p} \left( \frac{\diff p}{\diff t} \psi_i  
    - \frac{p}{3} (\bm{\nabla \cdot V}) \psi_i \right) -
    \frac{1}{\tau_{f,i}} \psi_i - \frac{1}{\tau_{r,i}} \psi_i.
\end{eqnarray}
The equation can either be solved analytically, utilizing various simplifying assumption \cite{Putze:2010zn,Maurin:2018rmm}, or fully numerically, 
as implemented in codes like \textsc{Galprop}~\cite{Strong:1998fr,Strong:2015zva}, \textsc{Dragon}~\cite{Evoli:2008dv,Evoli:2017vim} and \textsc{Picard}~\cite{Kissmann:2014sia}.

For a given primary CR species $i$, the equation has a source term $q_i(\bm{x},p)$ 
which is assumed to factorize into a space- and rigidity-dependent part, 
\begin{equation}
q_i(\bm{x},p) = q_i(r,z,R) = q_{0,i} \, q _{r,z}(r,z) \, q_{i,R}(R)\,,
\end{equation}
where $r$ and $z$ are cylindrical coordinates with respect to the Galactic center and $R$ denotes the rigidity. The rigidity dependence 
is taken to be a smoothly broken power law with break position $R_0$, and spectral indices $\gamma_{1,i}$ and $\gamma_{2,i}$ above and below the break, respectively,  while the smoothing is controlled by a parameter $s$:
\begin{eqnarray}
  \label{eqn::SourceTerm_2}
  q_R(R)     &=&   \left( \frac{R}{R_0} \right)^{-\gamma_1}
                  \left( \frac{R_0^{1/s}+R^{1/s}}
                              {2\,R_0^{1/s}                } \right)^{-s (\gamma_2-\gamma_1)}.
\end{eqnarray}
The spatial dependence of the source term is parametrized as
\begin{eqnarray}
  \label{eqn::SourceTerm_distribution}
  q_{r,z}(r,z)  = \left( \frac{r}{r_s} \right)^\alpha \exp \left( -\beta \, \frac{r-r_s}{r_s} \right) 
                                                    \exp \left( - \frac{|z|}{z_0} \right),
\end{eqnarray}
with parameters $\alpha = 0.5$, $\beta=2.2$, $r_s=8.5$~kpc, and $z_0=0.2$ kpc.\footnote{These are the default values in \textsc{Galprop v56}, which slightly differ from the values obtained from supernova remnants \cite{Case_SNR_Distribution_1998,Green:2015isa}. However, as pointed out in~\cite{Korsmeier:2016kha} the spatial dependence of the source term distribution has a negligible effect on the local CR fluxes.} 

Several processes contribute to CR propagation, in particular diffusion, reacceleration, convection and energy losses.
Spatial diffusion is assumed to be isotropic and homogeneous and is described by a rigidity-dependent diffusion coefficient $D_{xx}$.
We use a broken power law in rigidity in light of the recent measurements  by AMS-02 of the secondaries boron, lithium and beryllium~\cite{Aguilar:2018njt},  
which favors the interpretation of the observed CR break around $300$ GV as a break in diffusion rather than CR injection:  
\begin{eqnarray}
  \label{eqn::diffusionConstant}
  D_{xx} &=&
   \begin{cases} 
   		\beta D_{0} \left( \frac{R}{4 \, \mathrm{GV}} \right)^{\delta} &\text{if}\; R<R_1 \;\text{and}\\
   		\beta D_{0} \left( \frac{R_1}{4 \, \mathrm{GV}} \right)^{\delta} \left( \frac{R}{R_1} \right)^{\delta_2} &\text{otherwise}\,,
   	\end{cases}
\end{eqnarray}
where $\delta$ and $\delta_2$ are, respectively, the indices below and above the break at $R_1$,  $D_0$ is the overall normalization, and
$\beta=v/c$ the velocity of the CRs. 
The velocity of convective winds, $\bm{V}(\bm{x})$, is assumed to be constant and orthogonal to the Galactic plane, 
$\bm{V}(\bm{x})= {\rm sign}(z)\, v_{0,{\rm c}}\,{\bm e}_z$. 
Diffusive reacceleration is parametrized by the velocity $v_{\rm A}$ of Alfv\`en magnetic waves \cite{Ginzburg:1990sk,1994ApJ...431..705S}:
\begin{eqnarray}
  \label{eqn::DiffusivReaccelerationConstant}
  D_{pp} = \frac{4 \left(p \, v_\mathrm{A} \right)^2 }{3(2-\delta)(2+\delta)(4-\delta)\, \delta \, D_{xx}}.
\end{eqnarray}
This formula exhibits an ambiguity regarding the choice of $\delta$ when, as in our case, spatial diffusion
has a break. We choose to use a single $\delta$ at all rigidities, specifically 
the $\delta$  below the break $R_1$ introduced in  Eq.~\eqref{eqn::diffusionConstant}.
Finally, CR propagation is affected by energy losses, continuous, adiabatic and catastrophic,
for which we use the default \textsc{Galprop} implementation.
Explicitly, in Eq.~\eqref{eqn::propagationEquation} continuous energy losses are described by the term $\diff p/\diff t$, and catastrophic losses through fragmentation or decay are described by the respective lifetimes $\tau_f$ and $\tau_r$.
Catastrophic energy losses  provide a source term for secondary CRs (see Sec.~\ref{sec::xs} for details). 
Consequently, the propagation equations for the different CR species (primary and secondaries) constitute
a coupled system of differential equations.

\medskip

Beside secondary production by CR proton and helium interactions with the ISM,  
annihilation of DM in the Galaxy may lead to an additional source of antiprotons from the fragmentation and decay of the annihilation products
\cite{Bergstrom:1999jc,Donato:2003xg,Bringmann:2006im,Donato:2008jk,Fornengo:2013xda,Hooper:2014ysa,Pettorino:2014sua,Boudaud:2014qra,Cembranos:2014wza,Cirelli:2014lwa,Bringmann:2014lpa,Giesen:2015ufa,Evoli:2015vaa,Evoli:2011id,Reinert:2017aga}. 
The corresponding source term reads
\begin{eqnarray}
  \label{eqn::DM_source_term}
  q_{\bar{p}}^{(\mathrm{DM})}(\bm{x}, T) = 
  \frac{1}{2} \left( \frac{\rho(\bm{x})}{m_\mathrm{DM}}\right)^2  \sum_f \left\langle \sigma v \right\rangle_f \frac{\diff N^f_{\bar{p}}}{\diff T} ,
\end{eqnarray}
where $m_\mathrm{DM}$ and $\rho(\bm{x})$ are the DM mass and DM energy-density profile, respectively. The sum runs over
all DM annihilation channels $f$. 
The corresponding velocity averaged cross section and antiproton energy spectrum per annihilation are 
denoted by $\left\langle \sigma v \right\rangle_f$ and $\diff N^f_{\bar{p}}/\diff T $, respectively, where $T$ is the kinetic energy. 
Note that the factor $1/2$ corresponds to Majorana fermion DM\@. The energy spectra per annihilation at production,
${\diff N^f_{\bar{p}}}/{\diff T}$, depend on the DM mass, the kinematics of the annihilation process and details of the fragmentation and decay of the annihilation products. In this article, we consider the annihilation into bottom quarks, ${\rm DM\; DM} \to b\bar{b}$, for illustration. Note that the antiproton spectra of other hadronic channels exhibit similar shapes. Therefore, we expect this choice to be suitable to analyze a potential DM signal even though the true composition of annihilation channels may be very different. The corresponding best-fit DM mass may, however, differ for different annihilation channels, as discussed in~\cite{Cuoco:2017rxb}. We employ the spectra presented in~\cite{Cirelli:2010xx}. 

The DM density profile is still subject to sizable uncertainties. However, CR probes of DM do not exhibit a strong
sensitivity on the chosen profile~\cite{Cuoco:2017iax}. Within this work we use the Navarro-Frenk-White density profile~\cite{Navarro:1995iw},
with a characteristic halo radius as $r_\mathrm{h}=20\,$kpc and normalization 
so as to obtain a local DM density $\rho_\odot = 0.43\,$GeV/cm$^3$~\cite{Salucci:2010qr} at the 
solar position $r_ \odot = 8\,$kpc. 

\bigskip

We assume that CRs are in a steady state, which is a good approximation for nuclei. 
 The geometry of the Galaxy is approximated by a cylindrical box with an extension of $r=20$~kpc and $z=\pm z_\text{h}$. 
We solve the diffusion equation numerically with the \textsc{Galprop} code\footnote{http://galprop.stanford.edu/} \cite{Strong:1998fr,Strong:2015zva}.
\textsc{Galprop} solves the equation on a grid in the kinetic energy per nucleon and in the two spatial dimensions $r$ and $z$.
The radial and $z$ grid steps are chosen as $\Delta r=1\,$kpc and $\Delta z = 0.1\,$kpc, respectively. 
The grid in kinetic energy per nucleon is logarithmic between $1$ and $10^7\,$MeV with a step factor of $1.4$. 
We use versions \textsc{Galprop}~56.0.2870 and \textsc{Galtoollibs}~855\footnote{https://galprop.stanford.edu/download.php} as the basis and implement several custom modifications. Most importantly, first, we include the antiproton production cross sections from di Mauro \emph{et al.}~\cite{Mauro_Antiproton_Cross_Section_2014} and Winkler~\cite{Winkler:2017xor} (with updated parameters from Korsmeier \emph{et al.}~\cite{Korsmeier:2018gcy}). We allow for either using default cross sections from tables or the full parametrization of the Lorentz invariant cross sections. In the latter case the Lorentz transformation and angular integration is performed on the fly.
Secondly, we implemented the possibility of using a smoothly broken power law in the injection spectrum.

\bigskip
At low energies, CRs are deflected and decelerated by solar winds and the solar magnetic field. The strength  varies in a 22-year cycle and is, therefore, commonly referred to as solar modulation. The effect starts to be significant below few tens of GV and increases towards low energies. 
Solar modulation can be described by a propagation equation similar to the one of interstellar CR propagation but adjusted to the situation in the heliosphere. 
It can be solved numerically and compared to data~\cite{Kappl:2015hxv,Maccione:2012cu,Vittino:2017fuh,Boschini:2017gic}. 
Progress in the understanding of CR propagation in the heliosphere has been made recently~\cite{Tomassetti:2017hbe,Cholis:2015gna} 
especially thanks to
(i) the data of the Voyager~I probe which has left the heliosphere a few years ago and thus determines the CR fluxes before they are influenced by the solar effects,
and (ii) the data of PAMELA, which provided time dependent CR fluxes. Further progress is expected from the analysis
of the time-dependent CR fluxes released recently  by AMS-02~\cite{Aguilar:2018wmi,Aguilar:2018ons}.
Nonetheless, a detailed understanding is still missing at the moment. For this reason we resort to the commonly used force-field approximation~\cite{Fisk_SolarModulation_1976}, where the CR flux near Earth is calculated as
\begin{eqnarray}
	\label{eqn::solarModulation}
	\phi_{\oplus,i}(E_{\oplus,i}) &=& \frac{E_{\oplus,i}^2 - m_i^2}{E_{\text{LIS}, i}^2 - m_i^2} \phi_{\text{LIS}, i}(E_{\text{LIS}, i}) \,,\\
	E_{\oplus,i} &=& E_{\text{LIS}, i} - e|Z_i|\varphi_{{\rm SM},i}\,.
\end{eqnarray} 
Here, $e$ 
is the elementary charge, $Z_i$ and $m_i$ are the charge number and mass of the species $i$, respectively,  and $\varphi_{{\rm SM},i}$ 
is the corresponding solar modulation potential. The variables $E_{\text{LIS},}$ and $E_{\oplus,i}$ describe the total CR energy before and after solar modulation, respectively. To minimize the deviations of solar modulation from the force-field approximation
we  limit the analysis to rigidities larger than 5 GV.
In this respect various comments are in order:
\begin{itemize}
\item [(i)]
The recent publication by AMS-02 on time-dependent proton and helium fluxes~\cite{Aguilar:2018wmi}  
indicates that the  proton to helium ratio is in general constant, except for data taken after May 2015 at rigidities below 3 GV.
We will use proton and helium data before May 2015, and we can thus use the same modulation for the two species.

\item [(ii)]
It is well established that solar modulation is charge-sign dependent.
The AMS-02 measurements of the time-dependent proton and helium fluxes~\cite{Aguilar:2018wmi} as well as the electron and positron fluxes~\cite{Aguilar:2018ons}
provide a further confirmation of this effect.
To take this effect into account  we will use a different solar modulation potential for antiprotons as compared to $p$ and He, see Sec.~\ref{sec:setup}.

\item [(iii)]
The AMS-02 time-dependent data show explicitly a deviation from the force-field approximation. For example, the
proton spectra from different periods exhibit crossings,\footnote{For instance, the proton fluxes from 
November 2013 (Bartels rotation 2460) and March 2015 (Bartels rotation 2476) cross at $\sim 4$\,GV.}
an effect that is not possible in the force-field approximation.
This in particular is seen at low energies below $\sim5$ GV and during the maximum of solar activity.

\end{itemize}

We conclude that below 5~GV the force-field approximation starts to lose its reliability. Hence we discard data below 5~GV by default in our analysis. Note also that the data we use do not
include the solar maximum. Nonetheless, we also performed test fits with different lower cuts on the rigidity, still within the force-field approximation. 
From these fits we will further show that a cut of 5 GV is conservative from the point of view of DM searches.

%===================================================================
\section{Default setup}\label{sec:setup}
%===================================================================

Here we provide a short summary of the fit setup which, in general, is rather similar to the one used in~\cite{Korsmeier:2016kha,Cuoco:2016eej}. However, there are some important differences which we point out below.
We use  AMS-02 proton and helium fluxes \cite{Aguilar_AMS_Proton_2015, Aguilar_AMS_Helium_2015}, 
which both span the data period from May 2011 to November 2013,
and the AMS-02 antiproton to proton ratio \cite{Aguilar:2016kjl},
taken during the period May 2011 to May 2015.
Furthermore, we use  proton and helium data from Voyager~\cite{Stone_VOYAGER_CR_LIS_FLUX_2013} 
and, in some fits, complement with data from CREAM~\cite{Yoon_CREAM_CR_ProtonHelium_2011}.

In general, the likelihood for the CR fit is given by the product of the likelihoods of all experiments and CR species:
\begin{eqnarray}
  	\label{eqn::likelihood_CR}
	-2\,\log({{\cal L}_{{\rm CR}} }) =  \chi^2_{\rm CR} =  \sum\limits_{e,s} \sum\limits_{i,j} 	
							\left(\phi^{(e)}_{{ s},i}- \phi^{(\text{m})}_{ s,e} (R_i)\right) 
							\left(\left(\mathcal{V}^{\left(e,s\right)}\right)^{\!-1} \right)_{ij}
							\left(\phi^{(e)}_{s,j}- \phi^{(\text{m})}_{s,e} (R_j)\right)\,.
\end{eqnarray}
Here $\phi^{(e)}_{{ s}, i}$ is the flux measured by the experiment $e$ for the CR species $s$ 
at the rigidity $R_i$ and $\phi^{(\text{m})}_{s,e}$ is the corresponding \textsc{Galprop} model. 
The covariance matrix $\mathcal{V}^{\left( e, s\right)}$ describes the uncertainty of the flux measurement. In the 
default setup we assume uncorrelated uncertainties, 
$\mathcal{V}_{ij}^{\left( e, s\right)} = \delta_{ij} \left[\sigma^{(e)}_{{s},i}\right]^2$. 
We suppress the explicit dependence of $\phi^\text{(m)}$ on all the fit parameters.

Cosmic-ray propagation is described by a total of 15 (or 17, when including DM) parameters.
They are partly described in Sec.~\ref{sec:cr}, but, for convenience, we list them again below.
Six parameters are used to describe the injection spectrum of protons and helium, 
i.e., the slopes below and above the rigidity break, $\gamma_{1,p}$,$\gamma_{1}$, $\gamma_{2,p}$, $\gamma_{2}$,
the common rigidity break $R_0$ and a common smoothing parameter $s$.
Five more parameters describe propagation, i.e.,  the normalization $D_0$ and slope
$\delta$ of the diffusion coefficient,
the velocity of Alfv\`en magnetic waves, $v_A$, the convection velocity, $v_{0c}$, 
and the Galaxy's half-height, $z_\mathrm{h}$. 
In the default setup we limit the fit range of AMS-02 data from 5 to 300 GV.
A more detailed discussion and justification of these numbers is given further below.
We notice, however, that in this way we avoid fitting the two parameters describing 
diffusion above 300 GV, which are instead fixed to $R_1=300$ GV and $\delta_2=\delta - 0.12$.
Two further parameters, $m_{\rm DM}$ and  $\sv$, are used to parametrize the considered DM model 
when antiprotons from DM annihilation are included in the fit.
These 11 (13) parameters are scanned using \textsc{MultiNest}~\cite{Feroz_MultiNest_2008}.
For the \textsc{MultiNest} setup we use 500 live points, an enlargement factor \textsc{efr=0.7},
and a stopping criterion of \textsc{tol=0.1}. The final efficiency of the scan is typically around 7\%
and the number of likelihood evaluation around 200\,000.

The remaining four parameters are treated in a simplified way.
Two are the normalization of the proton and helium fluxes, $A_p$ and $A_\mathrm{He}$, respectively, 
and the other two are the solar modulation potential, $\varphi_{\text{SM,AMS-02},p,{\rm He}}$,
referring to $p$ and He as well as $\varphi_{\text{SM,AMS-02},\pb}$ referring to $\pb$.
We do not apply priors on $\varphi_{\text{SM,AMS-02},p,{\rm He}}$,
while for $\varphi_{\text{SM,AMS-02},\pb}$ we apply a very weak Gaussian prior,
i.e., we add to the main likelihood the term
$-2\,\log({{\cal L}_{{\rm SM}} }) = (\varphi_{\text{SM,AMS-02},p,{\rm He}}-\varphi_{\text{SM,AMS-02},\pb})^2 / \sigma_\varphi^2$
where $\sigma_\varphi = 100$ MV.
We profile over these four parameters on-the-fly at each \textsc{MultiNest} likelihood evaluation following~\cite{Rolke:2004mj}.
More precisely, for each evaluation in the fit within the 11- (13-) dimensional parameter space
the likelihood is maximized over the four remaining parameters. This maximization is performed
with \textsc{Minuit}~\cite{James:1975dr}.
To interpret the scan result we use a frequentist framework, and we build one- and two-dimensional
profile likelihoods in the different parameters, from which we derive contours which are shown in various figures in the following.  

\medskip
There is a subtlety in the treatment of solar modulation. As mentioned above, the AMS-02 $\pb/p$ ratio data -- and hence the proton data used in this ratio --
are taken from a different period than the AMS-02 $p$ and He data. 
Thus the two $p$ datasets should be modulated by a different solar modulation potential to be self-consistent.
An obvious improvement would be to use $p$ and $\pb/p$ datasets from the same time period, which are, however, not available 
in the AMS-02 publication \cite{Aguilar:2016kjl}.
Alternatively, the $\pb$ absolute data (which are available in \cite{Aguilar:2016kjl}) could be fitted instead of the  $\pb/p$ ratio. This procedure is, however,
suboptimal, because in the ratio some systematic uncertainty cancels out and thus
the $\pb$ data have a larger relative error than the $\pb/p$ ratio data. Moreover,
the use of the $\pb/p$ ratio data considerably simplifies the fit
since the ratio is
considerably less sensitive to the injection parameters than the 
absolute $\pb$ flux.
To resolve this issue we proceed as follows. We derive an ``effective''  $p$
spectrum from  $\pb/p$ ratio and $\pb$ flux from \cite{Aguilar:2016kjl} taking $\pb / (\pb/p)$.
We then divide the published $p$ spectrum \cite{Aguilar_AMS_Proton_2015} by this
effective $p$ spectrum. As expected, this ratio is consistent with 1 above $\sim40$ GV
and slowly decreases at low rigidities due to solar modulation, up to a maximum deviation of 5\%
at 1 GV\@.
As a function of rigidity this ratio is then approximated by a smooth log-parabola,
which in turn is used to multiply the published $p$ and He data 
\cite{Aguilar_AMS_Proton_2015, Aguilar_AMS_Helium_2015}
to create an effective $p$ and He data set that corresponds to the same period as the $\pb/p$ ratio.
During our study the AMS-02 publication~\cite{Aguilar:2018wmi} became available providing time-dependent $p$
and He fluxes. This provides us with an alternative possibility to derive the $p$ and He effective fluxes
corresponding to the period of the $\pb/p$ ratio, namely, averaging the given monthly fluxes over that period.
This second method is expected to be more robust. Nonetheless, we found that the effective
 $p$ and He fluxes build from the two methods perfectly agree, except below 3 GV
 where we find a difference of the order of 2\%, which is comparable with the error bars.
 Using these effective $p$ and He fluxes we can self-consistently use the modulation
 potential $\varphi_{\text{SM,AMS-02},p,{\rm He}}$ for the absolute fluxes of $p$
 and He and for the $p$ in the $\pb/p$ ratio, and the modulation potential
$\varphi_{\text{SM,AMS-02},\pb}$ for the $\pb$ in the $\pb/p$ ratio.
In the future, to avoid these issues, we recommend experimental
collaborations to periodically release global datasets including
the measurements of the fluxes of all species available and all referring
to the same time period.
 
As a default production cross section for the secondary antiprotons
we use the model from Winkler et  al.~\cite{Kappl:2014hha,Winkler:2017xor} with updated parameters from Korsmeier \emph{et al.}~\cite{Korsmeier:2018gcy} (referred to as param.~MW in the following). The results of an alternative 
cross-section parametrization by di Mauro \emph{et al.}~\cite{Mauro_Antiproton_Cross_Section_2014} (param.~MD)
and the effect of uncertainties of the cross-section parameters are discussed in detail in Sec.~\ref{sec::xs}.

%=====================
%    \                                           |
%      \                                         |
%        \                                       |
\begin{figure}[t!]
\setlength{\unitlength}{0.1\textwidth}
    \begin{picture}(9,4.3)
      \put(0,0){\includegraphics[width=0.92\textwidth]{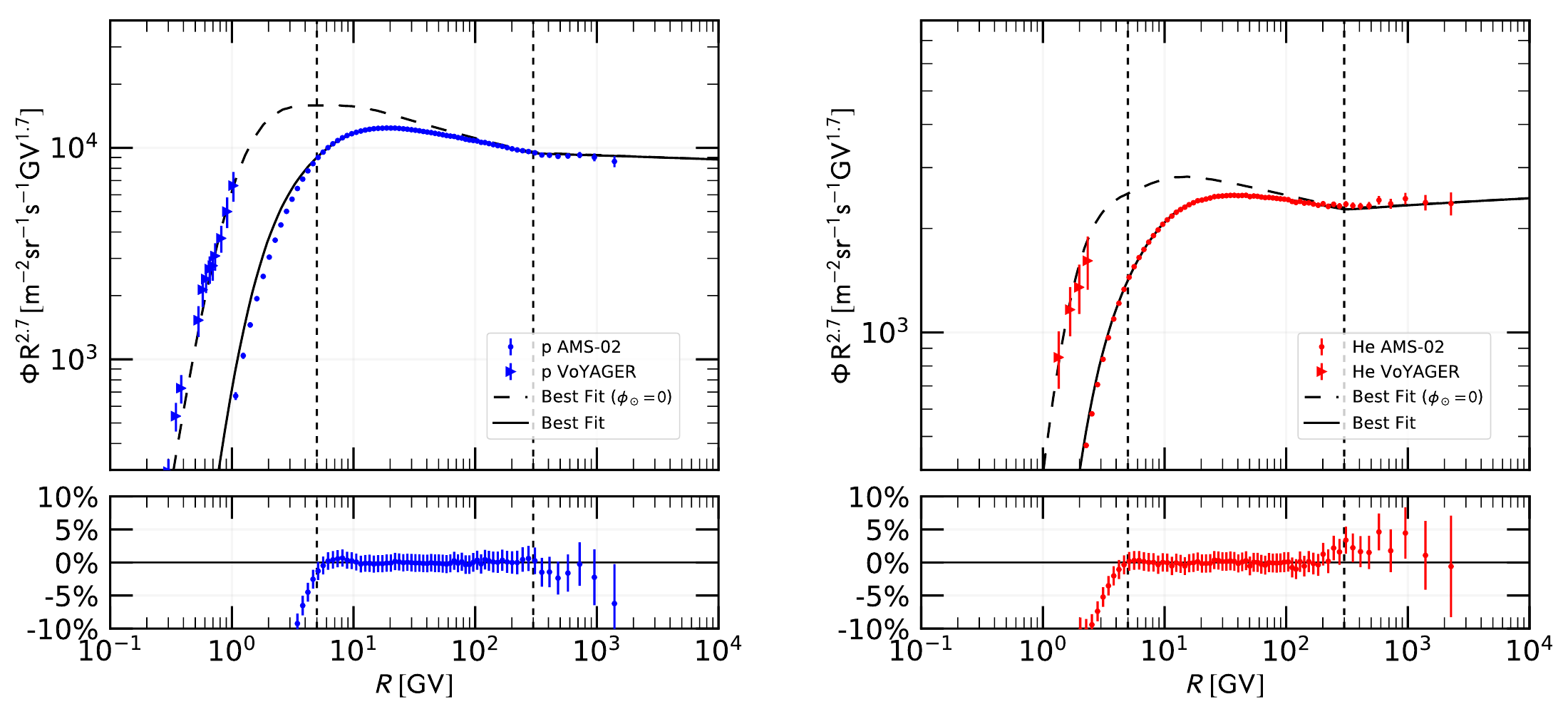}}     f
    \end{picture}
	\caption{Comparison of our best-fit proton and helium fluxes as a function of rigidity with AMS-02 and Voyager data. Both plots show the default setup without DM\@. In addition, we show the best fit before solar modulation ($\varphi_\odot=0$). The fit range is $R=(5\!-\!300)$\,GV (between the dotted lines).}
  \label{fig::pHe_spectra_standard_fit}
\end{figure}
%                                      \         |
%                                        \       |
%                                          \     |
%=====================

%=====================
%    \                                           |
%      \                                         |
%        \                                       |
\begin{figure}[t!]
\setlength{\unitlength}{0.1\textwidth}
    \begin{picture}(9.1,4.3)
      \put(0,0){\includegraphics[width=0.92\textwidth]{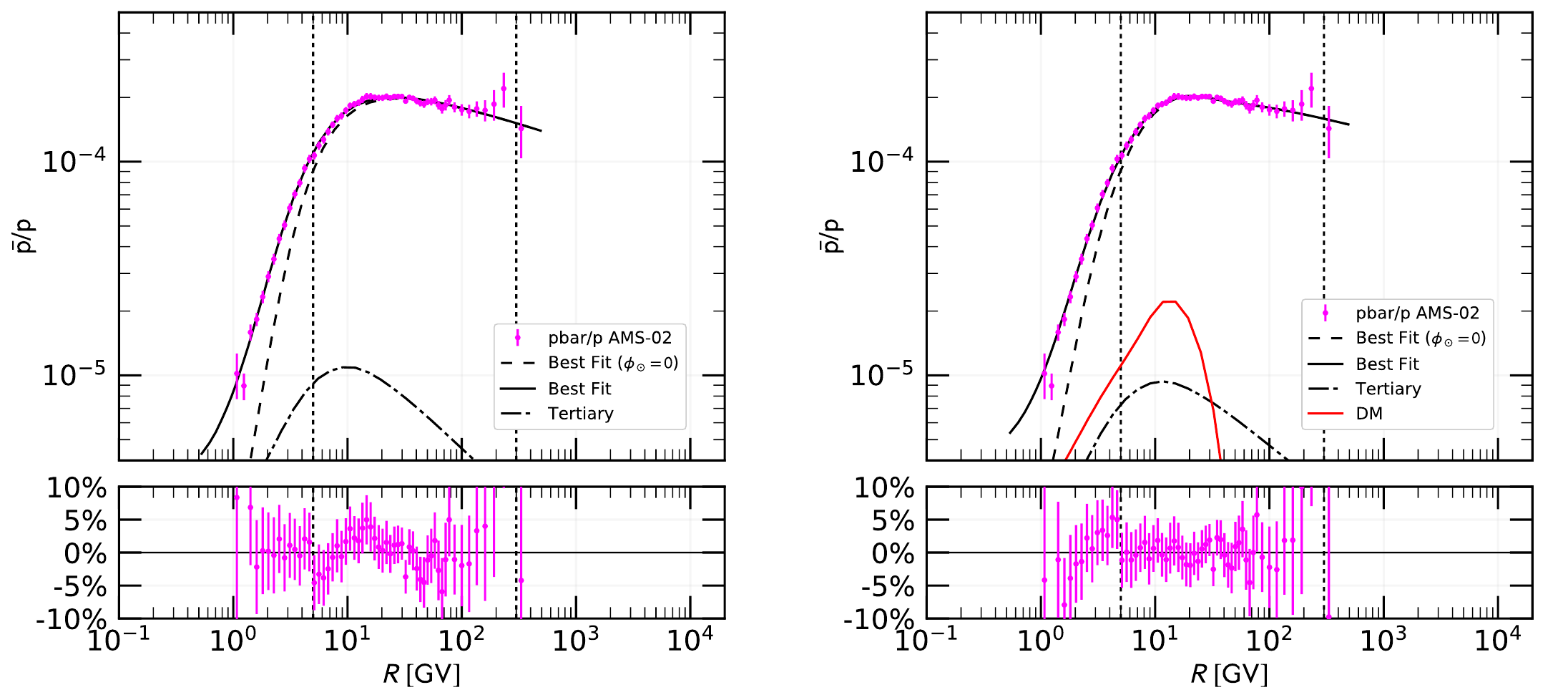}}     
    \end{picture}
	\caption{Comparison of our best-fit antiproton-over-proton ratio as a function of rigidity with AMS-02 data. The left plot shows the default setup without DM, while the plot in the right panel shows the corresponding setup with DM\@. In addition, we show the tertiary component, the DM component, and the best fit before solar modulation ($\varphi_\odot=0$). The fit range is $R=5$ to 300~GV (between the dotted lines).}
  \label{fig::pbar_spectra_standard_fits}
\end{figure}
%                                      \         |
%                                        \       |
%                                          \     |
%=====================

%=====================
%    \                                           |
%      \                                         |
%        \                                       |
\begin{figure}[t!]
	\setlength{\unitlength}{0.1\textwidth}
    \begin{picture}(10,9.2)
      \put( 0.00, 0){\includegraphics[width=0.92\textwidth]{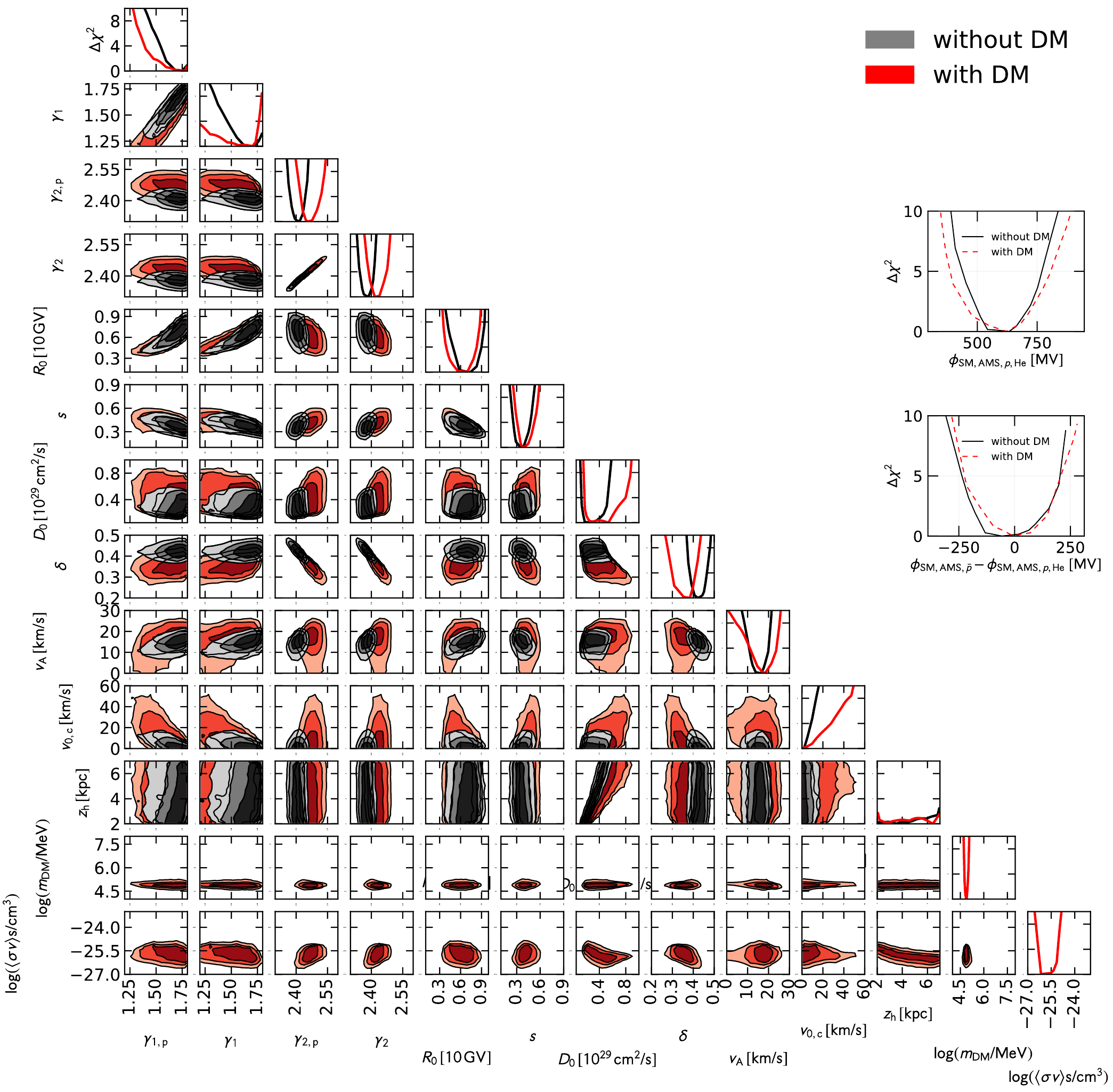}}     
    \end{picture}
	\caption{  Triangle plot with the fit parameters of the default fit which is the baseline for the following analyses. 
	           The black (red) contours show the 1$\sigma$ to 3$\sigma$ best-fit regions in the setup without (with) DM\@. 
	           On the diagonal the $\chi^2$ profiles are plotted for every fit parameter.
	           The two additional plots show the $\chi^2$ profiles of the solar modulation potential of AMS-02 $p$, He and 
	           its difference to the potential of $\pb$, respectively.
	}
  \label{fig::trinagle_standard_fits}
\end{figure}
%                                      \         |
%                                        \       |
%                                          \     |
%=====================

 \medskip
The results of the fit with this default setup, with and without DM, are shown in 
Figs.~\ref{fig::pHe_spectra_standard_fit}--\ref{fig::trinagle_standard_fits} and listed in the first two columns of Tab.~\ref{tab::bestfit_parameter1}.
Figure~\ref{fig::pHe_spectra_standard_fit} contains the comparison of the best-fit proton and helium spectra to data from AMS-02 and Voyager. The respective plots from the fit including DM look very similar. The residuals in the lower panels show a perfect agreement of the AMS-02 data with the \textsc{Galprop} model. However, they also already hint at a problem: The fluctuation of data points around the best fit is much smaller that the dominating systematic uncertainty. If this uncertainty is taken to be uncorrelated the fit results in a $\chi^2$ per degree of freedom (dof) much smaller than 1. We elaborate on possible correlation scenarios in more detail in Sec.~\ref{sec::corr}. 

Figure~\ref{fig::pbar_spectra_standard_fits} shows the best fit of the antiproton-to-proton ratio. Considering the fit regime from 5 to 300~GV there is a clearly visible improvement in the residuals, if DM is included in the fit. 
The significance in terms of $\chi^2$ difference between the fits excluding and including DM is $\Delta \chi^2\simeq12.7$ which formally corresponds to $3.1\sigma$. For the considered $b\bar b$ channel an annihilation cross section around $10^{-26}\,\text{cm}^3/s$ and a mass around 75 GeV provides the best fit.
As already found in~\cite{Cuoco:2016eej} we point out that the DM signature constitutes a spectral shape which is very different from the astrophysical secondary or tertiary components. 

Finally, we summarize the best-fit CR and DM parameters in the triangle plot displayed in Fig.~\ref{fig::trinagle_standard_fits}. Note that, compared to previous results in~\cite{Cuoco:2016eej} and due to the change of the standard cross section, the value of $\delta$ has increased to 0.42 and 0.38 in the case without and with DM, respectively. These values are in much better agreement with expectations from B/C data which points to 0.4--0.45~\cite{Genolini_CR_uncertainties_2015,Kappl:2015bqa}. The additional, embedded plots display the profiles for solar modulation. There is a small but nonsignificant preference for a slightly smaller solar modulation potential for antiprotons. 
From a theoretical point of view it is not exactly clear whether we expect a larger or smaller solar modulation potential for protons or antiprotons. The behavior is expected to depend on the polarity of the solar magnetic field. Since there is a change of polarity in 2013, which is in the middle of the period of the AMS-02 measurement of the antiproton-to-proton ratio, the exact situation is unclear. Future more detailed analyses of the recent monthly data by AMS-02 will provide a better understanding of this issue. 

\medskip
In the following we will discuss step by step the difference in our default setup with respect to the previous analysis performed in~\cite{Cuoco:2016eej}.
In particular there are three main points: We 
\begin{enumerate}
\item[(i)] removed data from CREAM, as it provides a source of tension with high-rigidity data from AMS-02
that is likely to stem from a systematic uncertainty in the normalization which were not properly addressed,
\item[(ii)] changed the $\pb$ production cross section from Tan{\&}Ng \cite{TanNg_AntiprotonParametrization_1983}
to the more recent param. MW,
\item[(iii)] introduced a separate solar modulation potential for $p$ and He, with respect to $\pb$ as described above.
\end{enumerate}
Besides the change of the best-fit propagation parameters (in particular $\delta$) mentioned above,
also the significance of the potential DM signal changes. In~\cite{Cuoco:2016eej} 
$\Delta \chi^2$ was found to be at the level of 25, corresponding to a significance of $\sim4.5\sigma$.
We performed several additional fits to trace the origin of the difference with respect to the current analysis. 
The $\Delta \chi^2$ changes 
\begin{enumerate}
\item[(i)] from 25 to 21, when removing CREAM data and fitting data only below 300~GV,
\item[(ii)] from 21 to 11, when changing the $\pb$ cross section to param.~MW,
\item[(iii)] and finally from 11 to 12.7, when introducing separate modulation potentials.
\end{enumerate}
This investigation shows that the cross-section parametrization has a potentially large impact on the DM significance.
We will investigate the robustness of the signal against cross-section uncertainties in the next section.

\medskip
In order to examine the compatibility of our new setup with CREAM data 
we perform one further fit. We include data from CREAM and allow for four additional fit parameters. 
First, we include the break position of the diffusion coefficient, $R_1$, and slope above this break, $\delta_2$,
which are sampled by \textsc{MultiNest}.
Secondly, we introduce two normalization parameters of the CREAM proton and helium data which
we leave free in order to take into account a possible systematic uncertainty (especially in the energy scale, 
which is degenerate with an uncertainty in the normalization for a pure power law the spectrum, as it is the case for the CREAM data in the limited energy range $\sim1$\,TeV--10\,TeV).
These nuisance parameters relieve the above-mentioned tension present when the data are used with the nominal normalization.
Going from the default fit to this extended setup with a total of 19 (21) parameters does not introduce 
significant complications since the two extra \textsc{MultiNest} parameters $R_1$ and $\delta_2$
are mainly determined by the data above 300 GV and are thus largely uncorrelated with the
rest of the parameter space, while the two normalization CREAM parameters are
profiled over similarly as described above.
The result of the fit is included in Tab.~\ref{tab::bestfit_parameter1} as the third and fourth columns.
It justifies our choice $R_1=300$ GV and $\delta_2=\delta - 0.12$ made above.
The DM significance is similar to the default setup and only increases slightly to  $\Delta\chi^2=15.1$, which
corresponds to $3.5\sigma$.

\medskip

Finally, as announced in Sec.~\ref{sec:cr} we briefly discuss the effect of taking into account data below 5\,GV
while still using the force-field approach to describe solar modulation. To this end we performed three additional fits following the above 
default setting (17 parameters) but with a rigidity cut at 3, 2 and 1\,GV.
The corresponding values for the $\chi^2$ are shown in Tab.~\ref{tab::chisq_below5}\@.
We observe a significant gradual worsening of the overall fit quality with decreasing rigidity, in particular between 2 and 1 GV\@.
This applies for both fits with and without DM\@.
It indicates that the model becomes less and less able to explain the data at lower energies, as expected from various effects like deviations from the force-field approximation, or, possibly, deviations from the simple scenario of convection-reacceleration.
The significance of a DM signal is maximal for a energy cut of 3 GV with a $\Delta \chi^2$ of around 19
and then decreases to $\Delta \chi^2\simeq11$ for a cut at 2 GV and $\Delta \chi^2\simeq0$ for a cut at 1 GV,
where, however, the overall fit quality is significantly worse as mentioned above.
The cut at 5 GV, hence, does not maximize the significance of a possible DM signal.
Note that the values of the other CR parameters do not change
significantly with the different rigidity cuts, while the errors have a slight improvement.
Thus, the estimation of the CR parameters seems robust
with respect to the variation of the lower rigidity cut.

%=====================================================================
%    \                                                                                        |
%      \                                                                                      |
%        \                                                                                    |
\begin{table}
    \caption{Fit quality for the best-fit parameter points without (second column) and with (third column) 
    DM for various choices of the rigidity cut. The last column shows the absolute 
    $\Delta \chi^2$ between the respective fits with and without DM\@. 
    }
  \centering
  \begin{tabular}{ c c c c c } \hline \hline
 & \multicolumn{2}{c}{$\chi^2/\text{ndf}$} &  &\\
  rigidity cut~[GV]  & excl.\,DM           & incl.\,DM & $\Delta \chi^2$ & \;DM significance \\
  \hline
  5 & $35.6/145 =0.245$  & $22.9/143 =0.160$   & 12.7 & $3.1\sigma$\\
  3 & $ 52.7/160 =0.329$ & $34.2/158 =0.216$   & 18.5 & $3.9\sigma$\\
  2 & $68.2/172 =0.396$  & $57.1/170 =0.336$   & 11.1 & $2.9\sigma$\\
  1 & $105.4/182 =0.579$\; & $105.6/180=0.586$ & -0.2 & -- \\ \hline\hline
  \end{tabular}
  \label{tab::chisq_below5}
\end{table}
%                                                                                  \         |
%                                                                                    \       |
%                                                                                      \     |
%=====================================================================

%===================================================================
\section{Antiproton cross sections}\label{sec::xs}
%===================================================================

\subsection{Introduction}

As we have seen in the previous section, the $\pb$ production cross section has an important impact on the fit and on the significance 
of the DM signal. In this section we thus go more into depth in the investigation of  this issue.
The CR antiprotons in our Galaxy are dominantly produced by the interaction of CR protons and helium with the ISM in the Galactic disk. The source term for the CR projectile nucleus $i$ and the ISM nuclei component $j$ is given by
\begin{eqnarray}
	\label{eqn::sourceTerm}
	q_{ij}({\bm x},\Tpbar) &=& \int\limits_{T_{\rm th}}^\infty \diff \Tproj \,\, 
                                    4\pi \,n_{\mathrm{ISM},j}({\bm x}) \, \phi_i  (\Tproj) \, \frac{\diff\sigma_{ij}}{\diff \Tpbar}(\Tproj , \Tpbar)\,.
\end{eqnarray}
Here $\phi_i$ is the CR flux, $n_{\mathrm{ISM},j}$ is the density of the ISM, and $\diff \sigma_{i,j}/\diff \Tpbar $ is the energy-differential cross section 
for antiproton production. The parameters $\Tproj$ and $\Tpbar$ denote the kinetic energy of the CR projectile and antiproton, respectively.
High-energy experiments measure the fully differential cross sections which are usually stated in the Lorentz-invariant form, $E_\pb\, \diff \sigma/\diff p_\pb^3$. There are two different strategies to extract the energy-differential cross section from the experimental data. On the one hand, Monte Carlo generators are tuned to the data and afterwards used to extract the required cross section \cite{Kachelriess:2015wpa,Lin:2016ezz}. However, at the moment these approaches lack consistency with data at either low or high energies, depending on the specific generator \cite{Donato:2017ywo}. On the other hand, an analytic parametrization of the Lorentz-invariant cross section is fitted to the experimental data. In this case, the energy-differential cross section is obtained by, first, performing a Lorentz transformation to the frame where the ISM component is at rest and, secondly, an angular integration~\cite{Mauro_Antiproton_Cross_Section_2014, Kappl:2014hha, Winkler:2017xor, Korsmeier:2018gcy}. This approach works reasonably well throughout the whole energy range of AMS-02, namely, from a rigidity of 1 to 400~GV. 
Therefore, we rely on the parametrization approach in the following. More details are given in~\cite{Donato:2017ywo, Korsmeier:2018gcy}.
We exploit the two parametrizations, param.~MW \cite{Winkler:2017xor} (used in the default setup above) and param.~MD \cite{Mauro_Antiproton_Cross_Section_2014}, for which we use the parameters updated to the most recent data from NA61 and LHCb as presented in~\cite{Korsmeier:2018gcy}. The uncertainty on the antiproton source term, solely due to cross sections, is at about 5\% above 
$T_\pb=5$~GeV and increases to 10\% below. To take this uncertainty properly into account we apply two different methods. 

\subsection{Covariance matrix method}

In the first method, we propagate the error of the cross-section parametrization to the flux of the CR antiprotons. 
This method was already suggested and applied in Ref.~\cite{Reinert:2017aga}. The procedure works as follows. We use the covariance matrix of the cross-section fits from Ref.~\cite{Korsmeier:2018gcy} and sample $N=1000$ random parameter combinations $k$ and the corresponding antiproton source terms $q^{(k)}_\pb$. 
From these we determine the covariance matrix of the relative source term $q^{(k)}_\pb(R_i)/q^{\text{(best\,fit)}}_\pb(R_i)$ at 
rigidities $R_i$ of AMS-02 
data points $i$. It is given by:
\begin{eqnarray}
  	\label{eqn::cov_XS_q}
	\mathcal{V}_{\text{XS},ij}^{(q_{\pb, \text{rel}})} = \frac{1}{N-1} \sum\limits_{k=1}^{N} \left(\frac{q^{(k)}_\pb(R_i)}{q^{(\text{best\,fit)}}_\pb(R_i)} -1\right)\left(\frac{q^{(k)}_\pb(R_j)}{q^{(\text{best\,fit)}}_\pb (R_j)} -1\right)\,.
\end{eqnarray}
(In formulas, tables and figures we abbreviate cross section with XS\@.)
We assume that the covariance matrices of the relative source term and of the relative flux are identical.\footnote{This is a good approximation since the relative uncertainty is invariant under propagation which is described by a linear differential equation.} In other words, the covariance matrix of the antiproton flux is given by
\begin{eqnarray}
  	\label{eqn::cov_XS_phi}
	\mathcal{V}_{\text{XS},ij}^{\left(\phi^\text{AMS-02}_{\pb/p}\right)} = \mathcal{V}_{\text{XS},ij}^{(q_{\pb,\text{rel}})} \, \phi^{(\text{AMS-02})}_{\pb/p,i} \,  \phi^{(\text{AMS-02})}_{\pb/p, j}\,. 
\end{eqnarray}
Here $\phi^{(\text{AMS-02})}_{\pb, i}$ is the antiproton flux measured by AMS-02 at the rigidity $R_i$.
Accordingly, for the log-likelihood of the antiproton data in our fit the covariance matrix is replaced by:
\begin{eqnarray}
  	\label{eqn::likelihood_pbar}
    \mathcal{V}_{ij}^{\left(\phi^\text{AMS-02}_{\pb/p}\right)} = 
        		\mathcal{V}_{\text{XS},ij}^{\left(\phi^\text{AMS-02}_{\pb/p}\right)} 
			  + \delta_{ij} \left[\sigma^{(\text{AMS-02})}_{\pb/p,i}\right]^{2}\,.
\end{eqnarray}
Obviously, this method has one weakness: Error propagation in terms of a covariance matrix assumes that the likelihood of the original cross-section parameters and the corresponding likelihood in the space of AMS-02 antiproton flux data points are well approximated by a multivariate Gaussian distribution. However, the true likelihood might be more complicated. 
Furthermore,  the method assumes that there is no correlation between the cross-section uncertainties and CR propagation uncertainties.
To take into account these shortcomings we thus consider a second method below.

\subsection{Joint fit method}\label{sec:jointxsfit}

In the second method we perform a joint fit of CRs and the antiproton production cross section. 
By simultaneously fitting the CR and cross-section parameters to CR fluxes and experimental cross-section data, the full likelihood is correctly taken into account. The price that we have to pay is an increase in the number of parameters for an already high-dimensional and hence time-consuming fit. 
In our default setup the CR fit contains 11--13  \textsc{MultiNest} parameters; the fit of cross section uses 7--10 free parameters. 
Thus, naively merging the fit would lead to ${\cal O} (20)$ free  \textsc{MultiNest} parameters, which is extremely challenging.
Therefore, we reduce the number of free parameters in the cross-section parametrization 
to those which affect the shape of the CR antiproton source term the most (as discussed further below) and fix the remaining parameters. 
We focus on param.~MW of the Lorentz-invariant cross section since the meaning of the single parameters is more obvious compared to param.~MD\@. The parametrization~MW for prompt antiproton production in proton-proton collisions depends on the center-of-mass energy $\sS$ of the collision, the energy of the antiproton divided by the maximal antiproton energy $\xR$, the transverse momentum of the antiproton $\pT$, and the six parameters $\mathcal{C}=\lbrace C_1...C_6\rbrace$:
\begin{eqnarray}
  \label{eqn::param_Winkler}
  E_\pb \frac{\diff^3 \sigma}{\diff p_\pb^3} (\sS, \xR, \pT) &=& \sigma_{\mathrm{in}}\, R_{\rm XS} \, C_1  (1-\xR)^{C_2} \, \left[ 1 + \frac{X}{\mathrm{GeV}}(m_\text{T} - m_p) \right]^{\frac{-1}{C_3 X}},
\end{eqnarray}
where $m_\text{T}=\sqrt{p_\text{T}^2 + m_p^2}$. The inelastic cross section $\sigma_{\mathrm{in}}$ of $pp$ scattering is defined in~\cite{Winkler:2017xor}.
The factor 
\begin{eqnarray}
  \label{eqn::param_Winkler_III}
   R_{\rm XS}  &=& 
   \begin{cases} 
     \left[ 1 + C_5 \left(10-\frac{\sS}{\mathrm{GeV}}\right)^5 \right]  \cdot \exp \left[ C_{6} \left( 10 - \frac{\sS}{\mathrm{GeV}} \right) (\xR - x_{R,\mathrm{min}} )^2  \right]& , \; \sS \leq 10\,\mathrm{GeV}   \\
     1 & ,\; \text{else} 
   \end{cases}
\end{eqnarray}
describes the scaling violation of the cross section at low $\sS$, and $X$ is defined by
\begin{eqnarray}
  \label{eqn::param_Winkler_IV}
  X &=& C_4 \log^2 \left( \frac{\sS}{4 m_p} \right).
\end{eqnarray}
For nonproton nuclei in the projectile CR or target ISM state we rescale the $pp$ cross section as described in~\cite{Korsmeier:2018gcy}. Furthermore, the total antiproton source term includes antiprotons produced by the decay of intermediate antineutrons or antihyperons. We apply the scalings from~\cite{Winkler:2017xor}.
The total likelihood for the joint fit is given by the product of the CR and cross-section likelihoods:
\begin{eqnarray}
	\label{eqn::likelihood_global}
	\log({\cal L}_\text{joint}) = \log({\cal L_{\rm CR,SM} }) + \log({\cal L_{\rm XS} })\,.
\end{eqnarray}
The procedure to fit the cross-section data follows~\cite{Korsmeier:2018gcy}. We fit to the same datasets 
(NA49~\cite{NA49_Anticic:2010_ppCollision}, NA61~\cite{Aduszkiewicz:2017sei}, Dekkers \emph{et al.}~\cite{Dekkers:1965zz}, NA49 ($p$C)~\cite{Baatar:2012fua}, LHCb ($p$He)~\cite{Graziani:2017}\footnote{During our analysis LHCb published the final analysis~\cite{Aaij:2018svt} of the cross section. They differ from the preliminary results by a scale factor of about 10\%. However, since we include a scale uncertainty of 10\% in our analysis, we do not expect a significant effect on the results.})
and use the same likelihood definition:
\begin{eqnarray}
	\label{eqn::likelihood_XS}
	-2\log({\cal L_{\rm XS}}) = \sum\limits_e  \sum\limits_i \left( \frac{{\omega_e}  \sigma_{\text{inv},i}^{(e)} -  \sigma_{\text{inv}}^{(m)}({\sS}_i,{\xR}_i,{\pT}_i)}{ {\omega_e} \sigma_{\sigma_{\text{inv},i}^{(e)}}} \right)^2 + \sum\limits_e  \left( \frac{1-\omega_e}{\sigma_{\omega_e}} \right)^2 \,.
\end{eqnarray}
As before, $e$ denotes experiments (this time of cross-section measurements) with data points $i$, while $m$ denotes the cross-section parametrization. The symbols $\sigma_\text{inv}$ and $\sigma_{\sigma_\text{inv}}$ represent the Lorentz-invariant cross section and its uncertainty, respectively.
We account for a scale uncertainty $\omega_e$ of each cross-section measurement $e$. It is constrained by a Gaussian prior, namely, the second term in Eq.~\eqref{eqn::likelihood_XS}. During the fit, these parameters are treated in a simplified way, similar to the normalizations and solar modulation parameters (cf.~Sec.~\ref{sec:setup}). For each \textsc{MultiNest} step the $\omega_e$s are profiled over by performing a \textsc{Minuit} fit. 
% 

%=====================
%    \                                           |
%      \                                         |
%        \                                       |
\begin{figure}[b!]
\setlength{\unitlength}{0.1\textwidth}
\begin{picture}(9.2,6.7)
      \put(0,0){\includegraphics[width=0.92\textwidth]{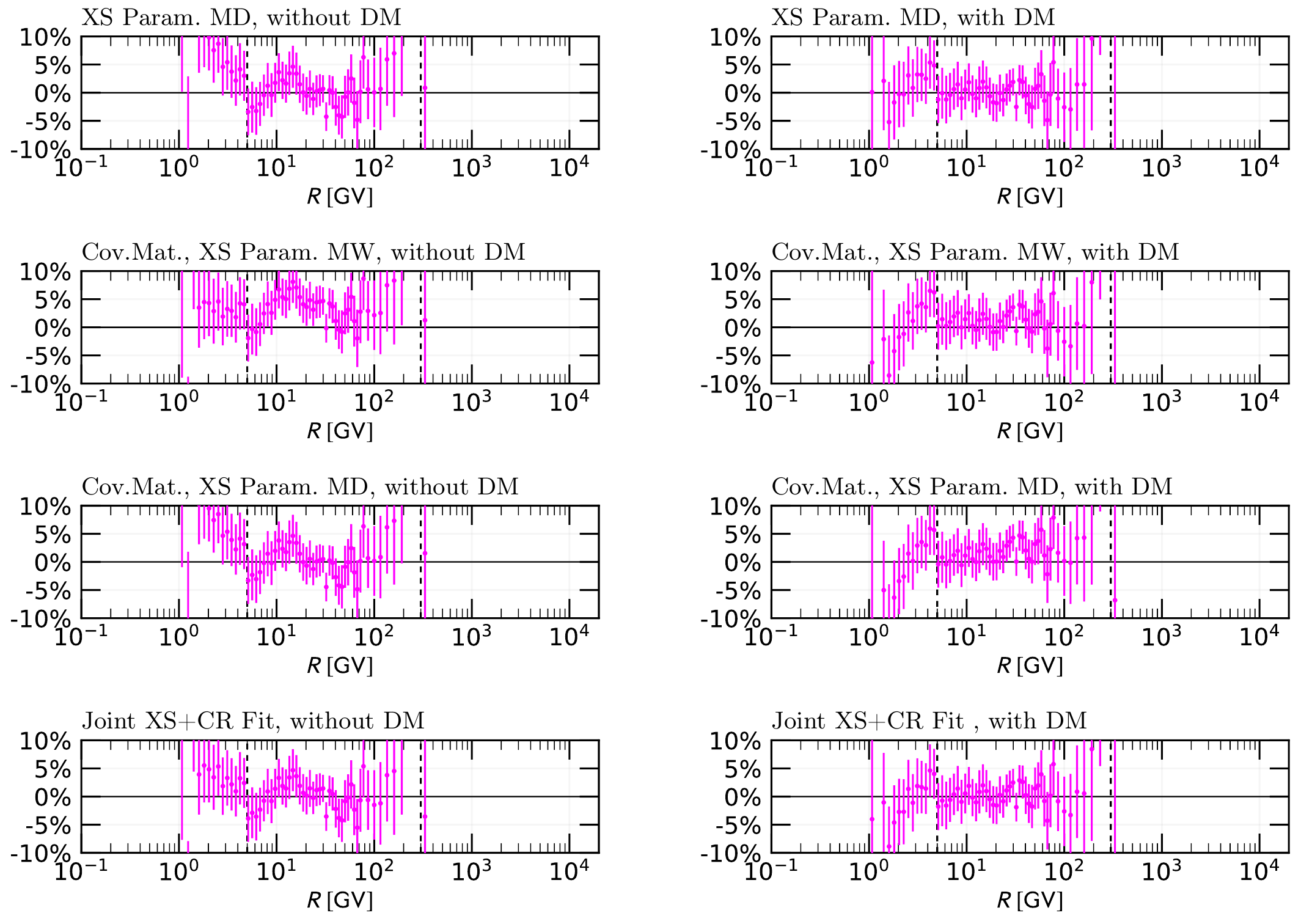}}     
    \end{picture}
	\caption{Residuals of the antiproton-over-proton ratio for different fit setups for the antiproton production cross section. The plot on the left-hand side originates from a fit setup without DM while the plot on the right-hand side is the corresponding setup including DM\@. From top to bottom the setups are changed compared to our default setup in the following way: (i) the parametrization of the antiproton production cross section is changed to param.~MD, (ii) cross-section uncertainties are treated effectively by means of a covariance matrix imposed on the antiproton data using the MW parameterization, (iii)  the cross-section uncertainties are treated effectively by means of a covariance matrix and the parametrization is changed to param.~MD, and (iv) in addition to the CR parameters we fit a selection of cross-section parameters simultaneously to CR and cross-section data (joint fit).
	  }
  \label{fig::pbar_residuals_xs}
\end{figure}
%                                      \         |
%                                        \       |
%                                          \     |
%=====================

 We now discuss our choice of parameters considered in the fit. The uncertainty of the antiproton production cross section has different origins. At high energies,  
above $\sS=10$~GeV, the shape of the cross-section data is constrained extremely well. The largest uncertainty is the normalization of the cross section. The origin of this uncertainty is the experimental difficulty to determine the luminosity better than a few percent.  At lower energies, data is more scarce and less precise. Furthermore, the theoretically motivated and experimentally confirmed concept of scaling invariance of the cross section is broken. Therefore, extrapolations are less trustworthy. 
We thus identify the parameters $C_1$ and $C_5$, $C_6$ as the most relevant for our purpose. They determine the normalization of the whole cross-section parametrization and the shape at low energies, respectively. In the following joint fit, we vary those parameters only, while all the other parameters are chosen to be fixed to the values from~\cite{Korsmeier:2018gcy}.
Note that adding these three parameters and the cross-section data to the global fit complicates the structure of the likelihood. Using the same configuration of \textsc{MultiNest} the total number of likelihood evaluations increases up to 500\,000. This is a clear disadvantage with respect to the approach with a covariance matrix which, instead, does not significantly complicate the fit.

%=====================
%    \                                           |
%      \                                         |
%        \                                       |
\begin{figure}[b!]
\setlength{\unitlength}{0.1\textwidth}
\begin{picture}(8,7.6)
      \put(0,-0.1){\includegraphics[width=0.8\textwidth]{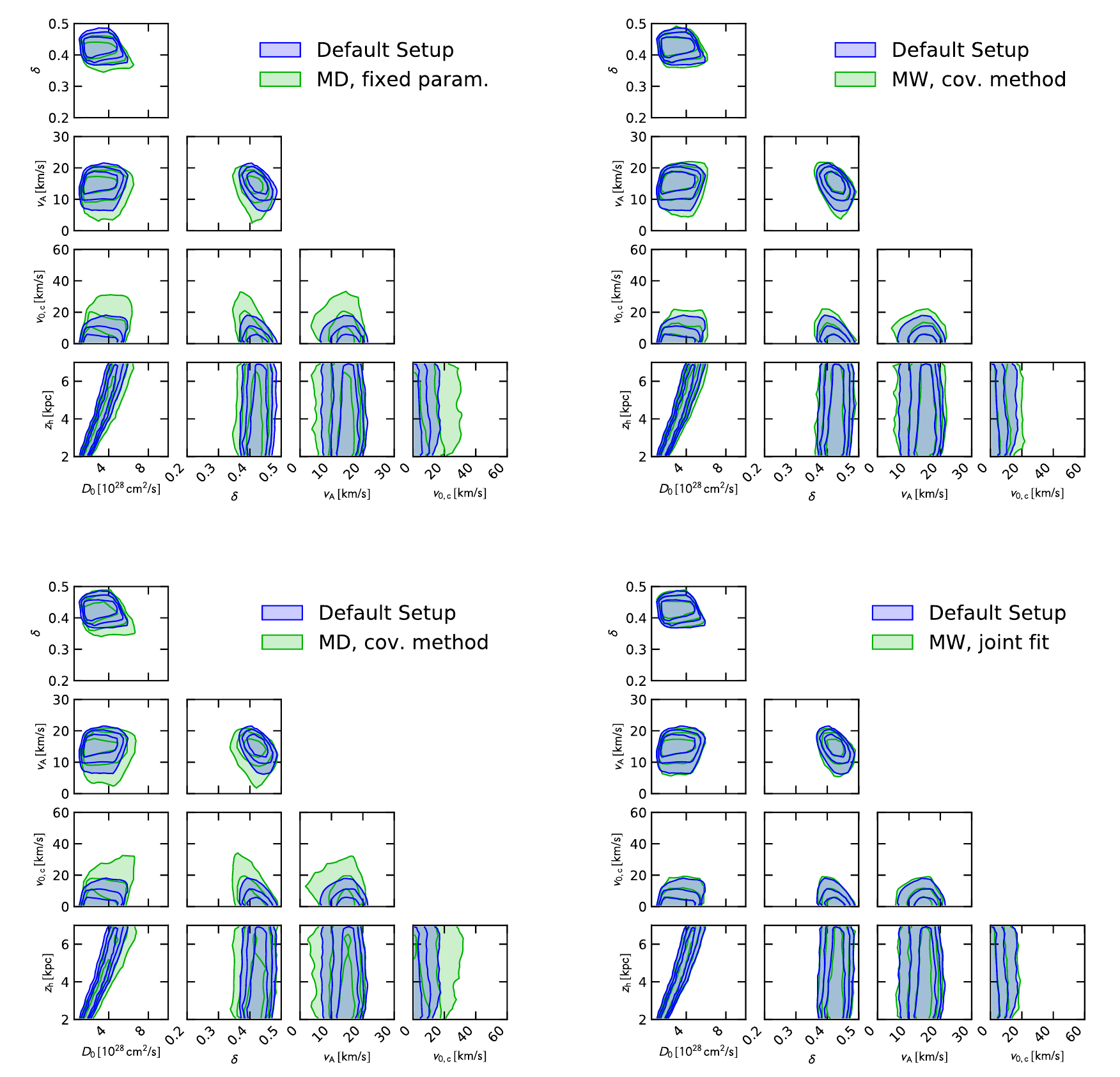}}     
    \end{picture}
	\caption{ Uncertainty contours (1--3$\sigma$) of the propagation parameters. 
	The triangles show the effect of changing the cross-section parametrization 
	from param.~MW to param.~MD (upper left), the effect of taking cross-section uncertainties 
	into account by a covariance matrix within param.~MW (upper right) and param.~MD (lower left), 
	and the effect of the joint fit (lower right).  
	}
  \label{fig::triangles_xs}
\end{figure}
%                                      \         |
%                                        \       |
%                                          \     |
%=====================

%=====================
%    \                                           |
%      \                                         |
%        \                                       |
\begin{figure}[t!]
    \setlength{\unitlength}{0.1\textwidth}
    \begin{picture}(9.8,2.8)
      \put(0,-0.1){\includegraphics[width=0.98\textwidth]{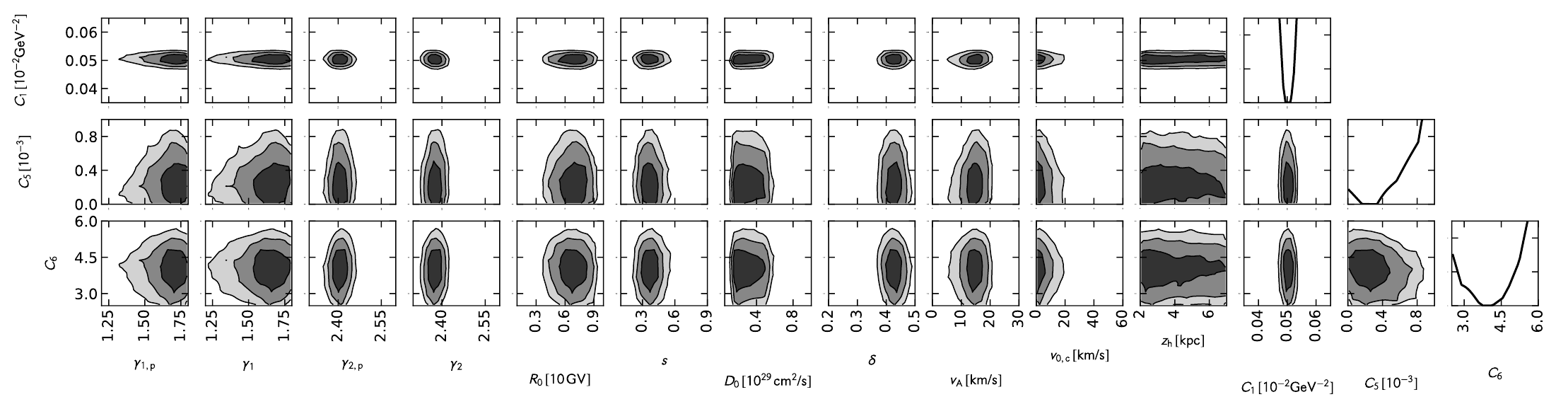}}     
    \end{picture}
	\caption{Correlation of the CR propagation parameters with the antiproton production cross-section parameters in the joint fit without DM\@. The black contours show the 1$\sigma$ to 3$\sigma$ region. The plot on the very right contain the $\chi^2$ profiles of the cross-section parameters. The y-axis ranges from $\Delta \chi^2 = 0$ to 10; cf.~Fig.~\ref{fig::chiSq_profiles_XS} for more details.  }
  \label{fig::triangle_XS_correlation}
\end{figure}
%                                      \         |
%                                        \       |
%                                          \     |
%=====================

%=====================
%    \                                           |
%      \                                         |
%        \                                       |
\begin{figure}[t!]
\setlength{\unitlength}{0.1\textwidth}
\begin{picture}(9.7,2.7)
      \put(0,-0.1){\includegraphics[width=0.97 \textwidth]{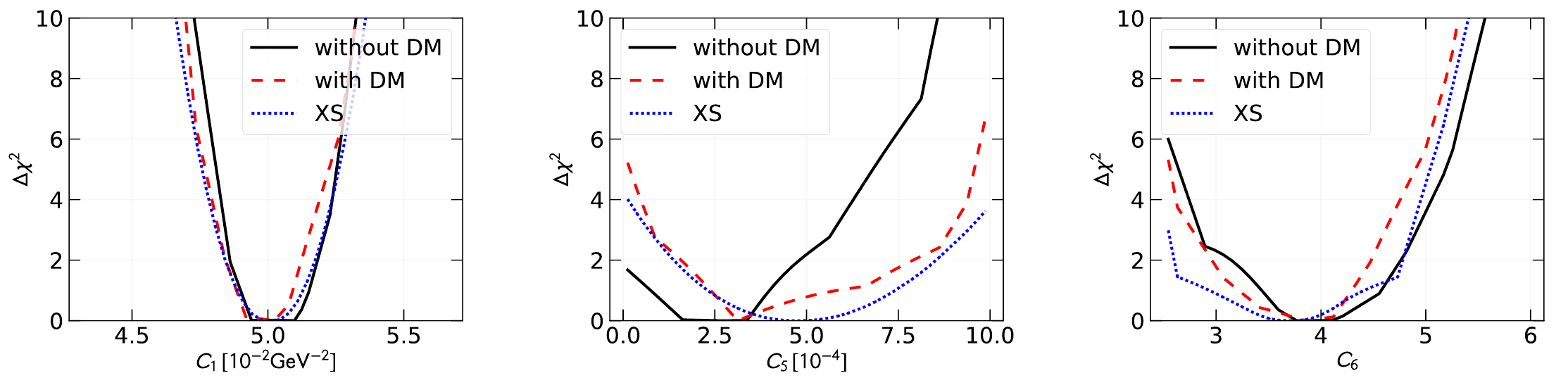}}     
    \end{picture}
	\caption{The three panels show the $\chi^2$ profiles of the cross-section parameter (cf.~Eq.~\eqref{eqn::param_Winkler}) included in the joint fit of CR and cross-section data. The black solid (red dashed) line shows the profile from the fit without (with) DM\@. For comparison we show the $\chi^2$ profile from cross-section data only (blue dotted line).}
  \label{fig::chiSq_profiles_XS}
\end{figure}
%                                      \         |
%                                        \       |
%                                          \     |
%=====================

\subsection{Results}

In total, we perform eight fits to study the effect of cross sections and their uncertainty on the results. These corresponds to 
four different setups with respect to the cross-section parametrization, while for each setup we perform one fit with and one without DM\@. The results of all fits are summarized in Tab.~\ref{tab::bestfit_parameter2}. The residuals of the antiproton spectra are shown in Fig.~\ref{fig::pbar_residuals_xs}. 
We start by discussing the effect on the fits without DM\@.
In the first setup, we change to cross-section param.~MD\@. As a result the fit quality improves marginally from $\chi^2_{\rm min} = 35.6$ to 34.2. The best-fit parameters are compatible and their uncertainty is very similar. 
In Fig.~\ref{fig::triangles_xs} we show the comparison of the fit contours with our default setup. The contours of this setup are slightly increased, in particular towards a larger $v_{0,{\rm c}}$, but they are not systematically shifted.
Then, we apply the covariance matrix method to both cross-section parametrizations. 
We find that the effect on the best-fit parameters and their uncertainties is negligible.  
It is, however, interesting to observe that in the residual plot of this fit there is a  systematic shift of all points towards larger values.
This is because in the energy range above 5 GV the cross-section covariance matrix $\mathcal{V}_\text{XS}^{\left(\phi^\text{AMS-02}_\pb\right)}$
mostly encodes the cross-section normalization uncertainty.

The results of the joint fit method are compatible with both the previous two fits, the one without cross-section uncertainties and the covariance matrix method. The residuals are mostly unchanged, and the parameter estimation and the corresponding uncertainty are similar.
Nevertheless, the joint fit provides very interesting insights allowing for further cross-checks. 
In particular, it allows us to investigate whether the CR parameters and the cross-section parametrization are correlated. 
In the extreme case, one might even imagine that the very precise CR data could constrain the cross-section parametrization.
Figure~\ref{fig::triangle_XS_correlation} displays the part of the parameter triangle which shows 
the correlation of the cross-section parameters with all CR parameters. 
We conclude that there is no significant correlation between the CR and cross-section fit parameter. 
Consequently, we expect that the cross-section parametrization is not affected by the CR data. 
This is confirmed in Fig.~\ref{fig::chiSq_profiles_XS}, which shows the $\Delta \chi^2$-profiles of the three cross-section parameters varied in the fit. 
We compare the profile of the total likelihood with the profile of a fit to only the cross-section data:
Both profiles agree well within their respective uncertainties. In fact, only $C_5$ is shifted to slightly lower values in the joint fit. 
This absence of correlation likely explains also why the covariance matrix method performs reasonably well 
and gives similar results.

\begin{figure}[t!]
\setlength{\unitlength}{0.1\textwidth}
\begin{picture}(10,3.5)
\put( 2.50, -0.2){\includegraphics[width=0.5\textwidth]{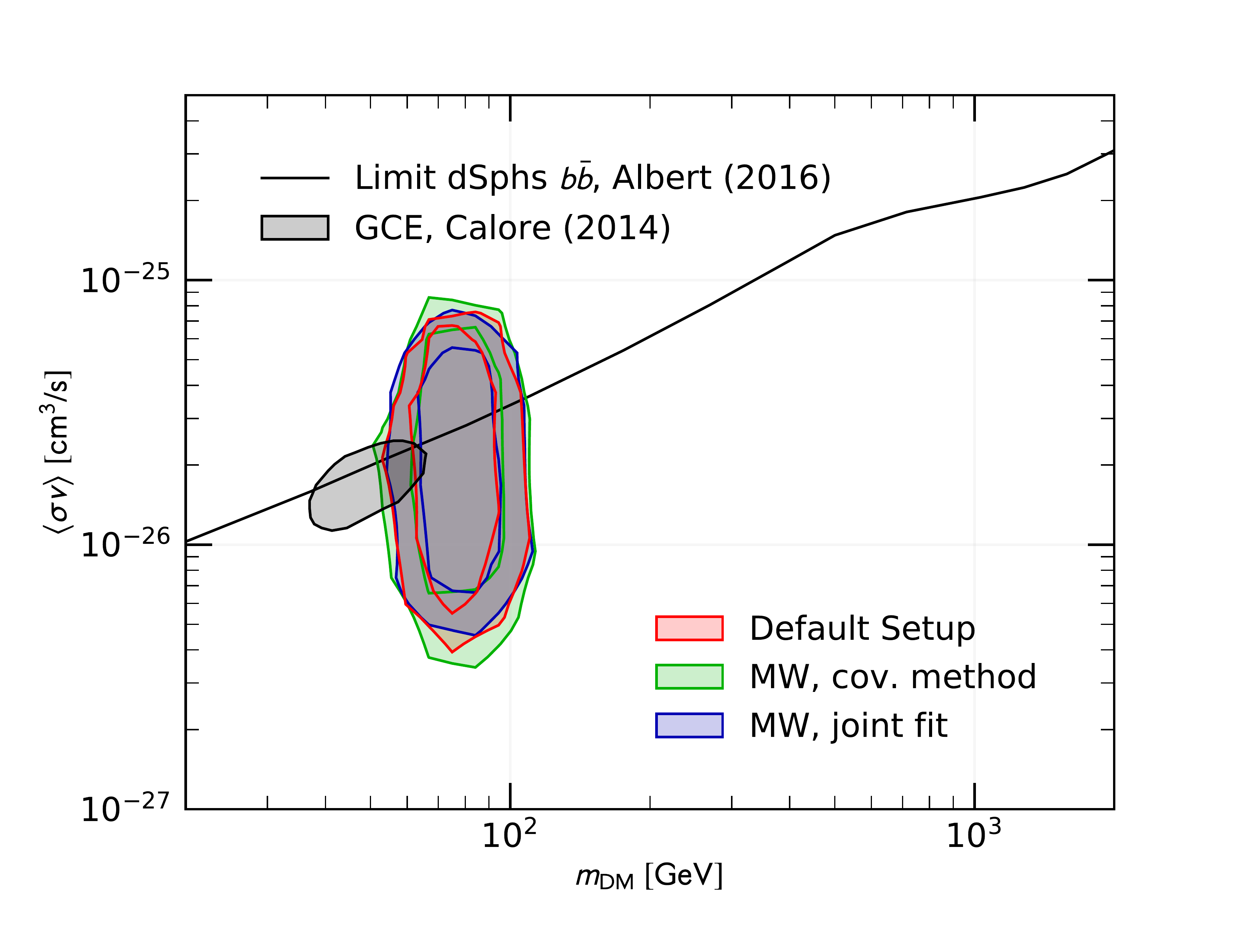}}
\end{picture}
	\caption{
	Contours of the 1$\sigma$ and 2$\sigma$ best-fit in the plane of DM mass and annihilation cross section.
	We overlay the result of the two different methods to treat cross section uncertainties, 
	the covariance matrix approach and the joint fit, with our default fit.  For comparison we show the 
	limit for the DM annihilation cross section derived from the observation of 
	dwarf spheroidal galaxies~\cite{Fermi-LAT:2016uux} and the $2\sigma$ best-fit region of the GCE~\cite{Calore:2014nla}.
	}
  \label{fig::DM_contours_XS}
\end{figure}

The impact of the cross-section uncertainty on the possible DM hint in the antiproton spectrum can be
understood looking at the residuals in Fig.~\ref{fig::pbar_residuals_xs} for the case of the fits with DM\@.
In all scenarios the flattening of the residuals is similar. In terms of $\chi^2$s 
the improvement of the fit with DM compared to the fit without DM was $\Delta \chi^2=12.7$ for the case of a fixed MW parametrization (default setup).
The covariance matrix and joint fit methods  decrease the  $\Delta \chi^2$ to 10.9 and 10.7, respectively, 
indicating that the evidence for DM is not strongly affected by the cross-section uncertainties.
Furthermore, the result of the best-fit DM mass and velocity-averaged annihilation cross section is
not strongly affected by the uncertainties. We show the comparison of the best-fit contours in 
Fig.~\ref{fig::DM_contours_XS}. 
For comparison we also show the cross-section limit derived from gamma-ray observations of 
dwarf spheroidal galaxies~\cite{Fermi-LAT:2016uux} and the best-fit region of the GCE~\cite{Calore:2014nla}
for the considered $b\bar b$ channel. All observations are compatible, in particular, since they are affected by	
astrophysical uncertainties in different ways providing additional freedom to alleviate a certain tension among them, see~\cite{Cuoco:2017rxb} for a detailed analysis of the subject.

Above, we have focused on fits and results where we exclude data below 5~GeV, since
as argued in the introduction, the results using data down to 1 GV are more prone to further systematic uncertainties, especially solar modulation. 
Nonetheless, it is interesting to have a look at the fit results including the low-energy data from a methodological point of view. 
The cross-section (shape) uncertainties are most severe at low energies, while at higher energies only the normalization is uncertain. 
Therefore, it is not very surprising that the results of both methods are very similar.
If, however, we include  data at low energies, the picture changes. 
We investigated how the best-fit parameters are affected by the two methods and find that both methods have a still small, 
but similar effect on the parameter space. Furthermore, we observe that the error contours of the covariance matrix method are a bit larger 
compared to the joint fit method, in other words, the former is more conservative.
We regard this as proof of concept: The covariance matrix method, which is easier to implement and less time consuming in the fit, 
is a reasonable approximation to the more complete joint fit method.

%=====================
%    \                                           |
%      \                                         |
%        \                                       |
%                                      \         |
%                                        \       |
%                                          \     |
%=====================f

The above results are somehow at odds with the results of~\cite{Reinert:2017aga}, where
flat $\pb$ residuals are achieved down to 1 GV and no significant preference for a DM signal (a global significance of $1.1\sigma$) was found.
The authors of this study use a covariance matrix method to account for the cross-section uncertainties.
They conclude that the inclusion of these uncertainties 
is the main reason why their analysis does not provide a hint for DM\@. 
Nonetheless, the results shown above indicate that the cross-section uncertainties do not
have such a strong impact.
An important difference is that in~\cite{Reinert:2017aga} only the $\pb$ spectrum is fitted,
with the source terms for $\pb$ being fixed using the observed $p$ and He spectra corrected for solar modulation.
This has the advantage that the injection parameters do not need to be fitted,
although it requires some assumption on how to extrapolate the observed local $p$ and He
spectra to the ones for the whole Galaxy needed for the secondary source terms.  
Instead, in our approach $\pb$, $p$ and He are fitted simultaneously and we include $p$ and He injection
parameters in the fit. 
Fitting the $p$ and He spectra provides extra constraints on the propagation with respect
to fitting $\pb$ only.
For example, it is well known, e.g.,~\cite{Moskalenko:2001ya},  that strong reacceleration produces a low-energy ($\lesssim 10$ GeV) bump
in the $p$ spectrum, which is not observed. The $p$ spectrum, thus, provides strong constraints
on the amount of reacceleration, although this is, in part, degenerate with the break in the injection~\cite{Moskalenko:2001ya}.
We thus suspect that in~\cite{Reinert:2017aga} it is possible to accommodate the secondary $\pb$ spectrum, 
while this is not possible anymore when constraints from $p$ and He are included as it is the case in our analysis.
Further differences concern a different treatment  of reacceleration 
(which in~\cite{Reinert:2017aga} is confined to the Galactic disk only, while it is uniform over the whole diffusion region in our case),  
adiabatic energy losses from convection and a two-dimensional source term distribution used in our analysis. 
Therefore a direct comparison is not easily achievable and would require a substantial 
modification of our setup, which is left for future work.

%===================================================================
\section{AMS-02 correlations}\label{sec::corr}
%===================================================================

With the era of space-based CR detectors  the statistics and quality of collected data have significantly increased.
This also means that the relative weight of systematic uncertainties with respect to the statistical error has become
more important. For example, the error budget of the measured proton and helium spectra
is now completely dominated by systematics in most of the energy range.
The question of how to assess and treat these uncertainties in a statistically correct way has thus become more pressing.
The commonly used strategy is to add statistic and systematic uncertainties in quadrature
and consider the new errors uncorrelated in energy.
This, as we have seen in the previous sections, typically results in quite low $\chi^2$ values of the fit to the data.
Table~\ref{tab::bestfit_parameter1} shows that the typical values of the $\chi^2/\text{dof}$ are of the order 0.06--0.1 for proton and helium and 0.25--0.50 for the antiproton-over-proton ratio, whereas for a consistent treatment of all uncertainties one would expect  a $\chi^2/\text{dof}$ close to 1.
We conclude that either the systematic uncertainties of AMS-02 are overestimated or that there is a sizable correlation of the systematic uncertainty which is not correctly taken into account. 
The fact that our $\chi^2$ values are significantly smaller than expected has potentially problematic consequences: 
First, we cannot say anything about the goodness of fit and, secondly, 
the uncertainties deduced for our fit parameters could be affected. 
Thirdly, the significance of the potential DM signal may depend on the correlations in the systematic uncertainties.
A proper assessment of this problem would require the experimental collaborations to provide the covariance matrix of the
data points based on the knowledge of the detector. 
Since this is not available at the moment, our goal below is to use a simple approach and gain an approximate understanding of the effects of neglecting correlations.

\subsection{Methodology}

We follow two different strategies to answer the above questions. 
In the first one, we simply set the systematic uncertainty to 1\% of the flux or flux ratio before adding it to the statistical uncertainty in squares.\footnote{A similar approach was taken in \cite{Cavasonza:2016qem}.} 
The typical systematic uncertainty stated in the AMS-02 publications of proton, helium, and antiproton fluxes is on the order of a few percent. 
The remaining systematic uncertainty is thus assumed to be fully correlated among the rigidity bins, i.e.~equivalent to an overall normalization uncertainty.
Since normalizations are already profiled over in our fit setup, this kind of uncertainty is already taken into account and does not need to be included again as an uncertainty in the data. 
Thus, in practice, this approach implies a significant reduction of the uncertainty with a potentially equally significant effect. 

The above approach, however, does not address the question about the presence of shorter-range correlations among the data points.
The most complete approach to the problem would be to start from the knowledge of the detector and model
the systematic uncertainties which contribute to the total error budget.
In the AMS-02 publications the various contributing systematics are listed. They contain the acceptance uncertainty,
trigger uncertainty, rigidity scale uncertainty, and uncertainty from energy unfolding.
By studying the single systematic effects and modeling them, it would be possible to build the covariance
matrix of the data. Unfortunately, again, this requires an inside knowledge of the detector which is not publicly available.
We thus resort, in the following, to a simpler, data-driven approach which is expected to model
the effect of correlated uncertainties reasonably well.
With this approach we aim at determining the systematic uncertainty and a possible correlation between data points.
The focus of this strategy, in particular, is on the study of a possible short-range (in rigidity) correlation component.
This component is potentially more critical for our analysis since it can affect the significance of sharp features like
the ones expected from DM\@. On the contrary, as argued above, long-range correlations
are basically equivalent to a normalization uncertainty and have a small impact.  
As a first step, we thus split the covariance matrix of the CR datasets into a sum of three parts:
\begin{eqnarray}
\label{eqn::covariance_matrix}
{\cal V} = {\cal V}_\text{stat} + {\cal V}_\text{short} +  {\cal V}_{\text{long}}\,.
\end{eqnarray}
Here, the first part ${\cal V}_\text{stat}$ contains all statistical, i.e.\ uncorrelated, uncertainties. 
In other words, the entries of the correlation matrix for the $i$-th and $j$-th data point are given by  ${\cal V}_{\text{stat},ij} = \delta_{ij} (\sigma_{\text{stat},i})^2$.  
The statistical error alone is generally available in the corresponding publications.
The third part ${\cal V}_{\text{long}}$ describes the long-range correlations. 
Typical examples which would fall into this kind of uncertainty are normalization or tilts of the whole data set. 
The second part  is ${\cal V}_\text{short}$ which describes the correlation of up to a few neighboring points. 
Our ansatz is
\begin{eqnarray}
\label{eqn::ansatz_Vshort}
{\cal V}_{\text{short},ij}=\exp\left(-\frac{|i-j|^\alpha}{{\ell_\text{corr}}^\alpha}\right)f^2\sigma_{\text{sys},i} \sigma_{\text{sys},j}\,,
\end{eqnarray}
where the three parameters ${\ell_\text{corr}}$, $f$, and $\alpha$ describe the correlation length (in terms of the distance in rigidity-bins),\footnote{The ansatz of Eq.~\eqref{eqn::ansatz_Vshort} assumes that the correlation length does not depend on the rigidity bin. We notice that the rigidity binning chosen by the AMS-02 experiment is inversely proportional to the energy resolution of the instrument. So, for example, an uncertainty in the energy unfolding is expected to be described by this kind of the covariance matrix.}
the fraction of the systematic uncertainty which is correlated, and the shape of the correlation matrix, respectively.  
Our goal is to determine the three parameters from the data themselves.
In practice, we regard the data as a realization of the true covariance matrix,
and we try to reconstruct them assuming the above parametrization, using
standard statistical inference.
To this end, besides the covariance matrix, we also need a model of the \emph{true} energy spectrum,
which we take as a smooth multiply broken power law with three breaks for the antiproton-to-proton ratio and the helium flux, and four breaks for proton flux. The parametrization is similar to Eq.~\eqref{eqn::SourceTerm_2} but extends to a higher number of breaks. 
Then, in principle, our inference should proceed with a fit of the covariance matrix parameters 
and spectrum parameters together to the observed data.
Here, instead, we apply a simplified two-step approach.
In the first step, we fit the smoothly broken power law to the data using as $\chi^2$
the full systematic uncertainties assumed to be uncorrelated.
In practice, this step is equivalent to absorbing the long-range correlated uncertainties into the smooth spectrum, and it is insensitive to the exact errors used to define
 the $\chi^2$.
In the second step we thus fix the smooth energy spectrum to the one derived in the above step,
and we use the residual with respect to the data points,  ${\bm x}$, to constrain the covariance
matrix. 
The log-likelihood for the parameters ${\ell_\text{corr}}$, $f$, and $\alpha$ is given by
\begin{eqnarray}
\label{eqn::likelihood_Vshort}
- \log({\cal L}) = \frac{1}{2} \log\left( \det\left[  {\cal V}_\text{stat} + {\cal V}_\text{short}({\ell_\text{corr}} , f, \alpha) \right] \right) + \frac{1}{2} {\bm x} \cdot \left[  {\cal V}_\text{stat} + {\cal V}_\text{short}({\ell_\text{corr}} , f, \alpha) \right]^{-1} \cdot {\bm x}  + \text{const}\,,
\end{eqnarray}
where the first term comes from the normalization of the multivariate Gaussian (in the data) which we use as likelihood.
We checked with a toy Monte Carlo that the method is self-consistent. 
To this end, in the Monte Carlo  we choose  benchmarks values of  ${\ell_\text{corr}}$, $f$, and $\alpha$,
i.e., a benchmark covariance matrix, and from that we draw random values of ${\bm x}$.
We then verify that from the above likelihood we correctly reconstruct the input values
of ${\ell_\text{corr}}$, $f$, and $\alpha$ within uncertainties.

%=====================
%    \                                           |
%      \                                         |
%        \                                       |
\begin{figure}[b!]
\setlength{\unitlength}{0.1\textwidth}
\begin{picture}(9.8,3.1)
      \put(0,-0.1){\includegraphics[width=0.98\textwidth]{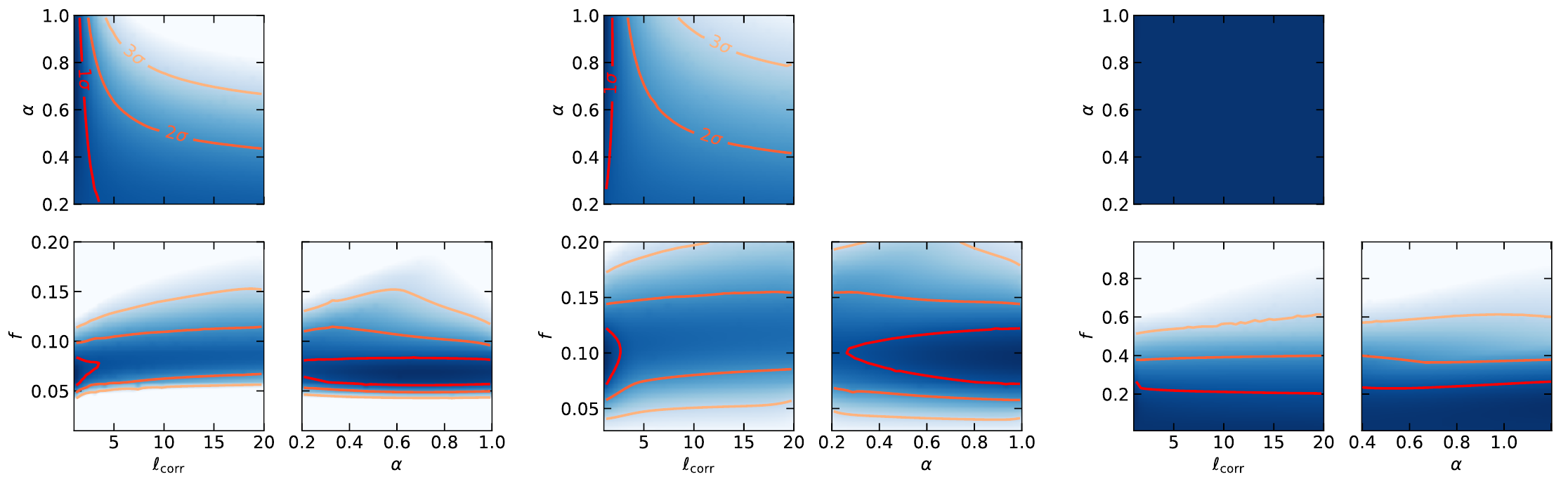}}     
    \end{picture}
	\caption{Triangle plots for the fit of a covariance matrix to proton, helium, and antiproton-over-proton data.
 }
  \label{fig::corr_datadriven}
\end{figure}
%                                      \         |
%                                        \       |
%                                          \     |
%=====================

%=====================================================================
%    \                                                                                        |
%      \                                                                                      |
%        \  
\begin{table}[b!]
\caption{Covariance matrices for the different benchmark scenarios. In the case of proton and helium, we maximize the log-likelihood in Eq.~\eqref{eqn::likelihood_Vshort} at fixed $\ell_\mathrm{corr}$ to determine $f$ and $\alpha$, while, to be conservative, we fix $f_\pb$ to 0.3 and $\alpha_\pb$ to 1.}
\label{tab::corr_benchmarks}
\begin{tabular}{c @{\hspace{20px}} c c c c c @{\hspace{20px}} c c c c c @{\hspace{20px}} c c c c c}
\hline \hline
                      & &  \multicolumn{3}{c}{$\bm{\ell_\text{corr}=0}$}  &&& \multicolumn{3}{c}{$\bm{\ell_\text{corr}=5}$}  &&& \multicolumn{3}{c}{$\bm{\ell_\text{corr}=10}$}  \\ \hline
   dataset           & &    $p$    &    He   &  $\pb/p$  &&&    $p$    &    He   &  $\pb/p$  &&&    $p$    &    He   &  $\pb/p$   \\   
   $f$                & &    0.062  &  0.080  &  0.30     &&&    0.079  &   0.103 &  0.30     &&&    0.082  &   0.101 &  0.30      \\
   $\alpha$           & &    0.63   &  0.81   &  1.00     &&&    0.20   &   0.21  &  1.00     &&&    0.20   &   0.20  &  1.00       \\ \hline \hline
	
\end{tabular}
\end{table}
%                                                                                  \         |
%                                                                                    \       |
%                                                                                      \     |
%=====================================================================

%=====================
%    \                                           |
%      \                                         |
%        \                                       |
\begin{figure}[b]
\setlength{\unitlength}{0.1\textwidth}
\begin{picture}(9.2,6.5)
      \put(0,0){\includegraphics[width=0.92\textwidth]{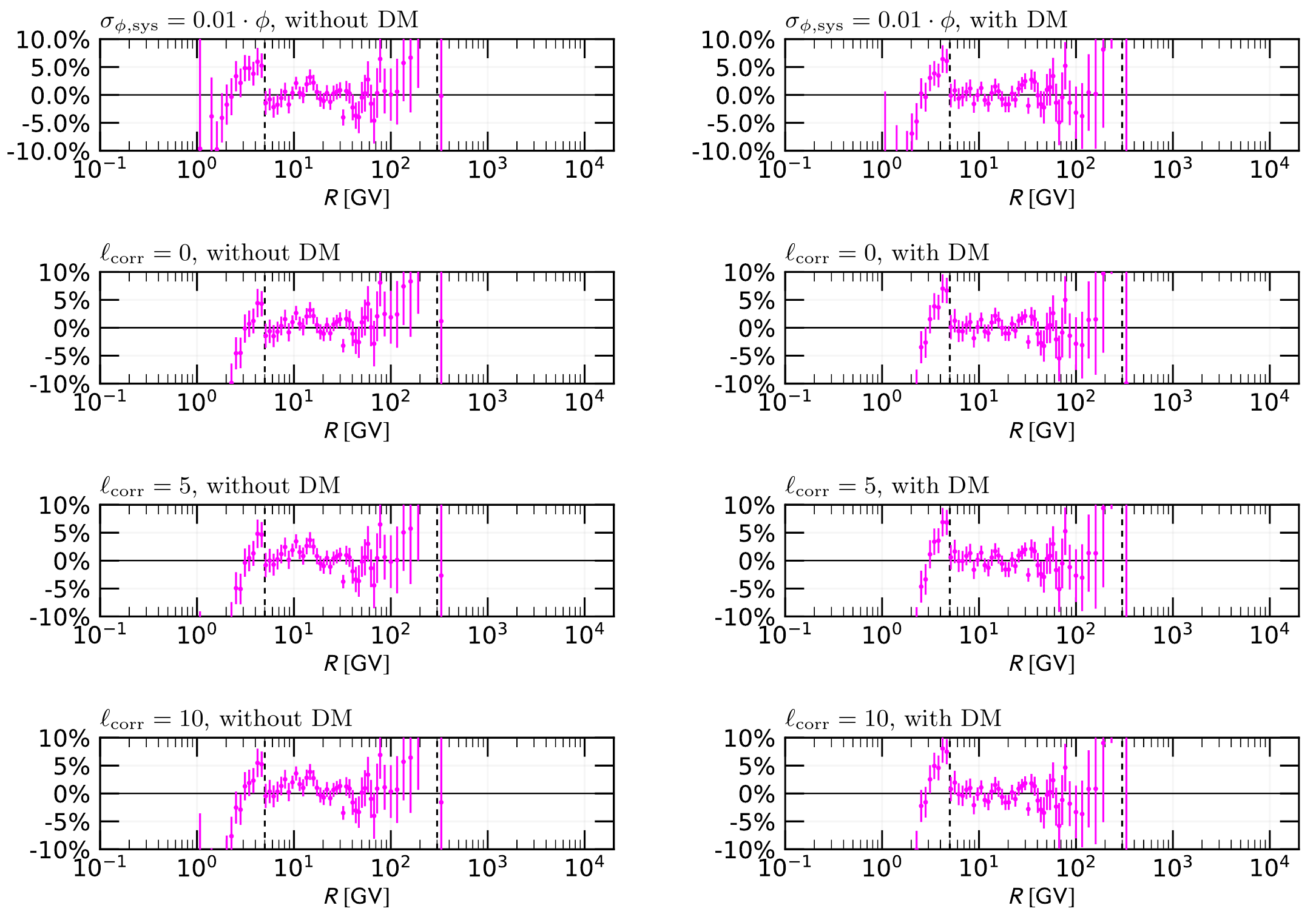}}     
    \end{picture}
	\caption{Residuals of the antiproton-over-proton ratio for different fit setups for different treatment of systematic uncertainties and its correlation. The plot on the left-hand side originates from a fit setup without DM while the plot on the right-hand side is the corresponding setup including DM\@. From top to bottom the setups are changed compared to our default setup in the following way: i) the systematic uncertainty is taken to be uncorrelated 1\% of the fluxes (or ratio), ii) data-driven correlation approach with a correlation length of $\ell_{\text{corr}}=0$, iii) $\ell_{\text{corr}}=5$, and iv) $\ell_{\text{corr}}=10$. }
  \label{fig::pbar_residuals_corr}
\end{figure}
%                                      \         |
%                                        \       |
%                                          \     |
%=====================

%=====================
%    \                                           |
%      \                                         |
%        \                                       |
\begin{figure}[b!]
\setlength{\unitlength}{0.1\textwidth}
\begin{picture}(8,7.9)
      \put(0,-0.1){\includegraphics[width=0.8\textwidth]{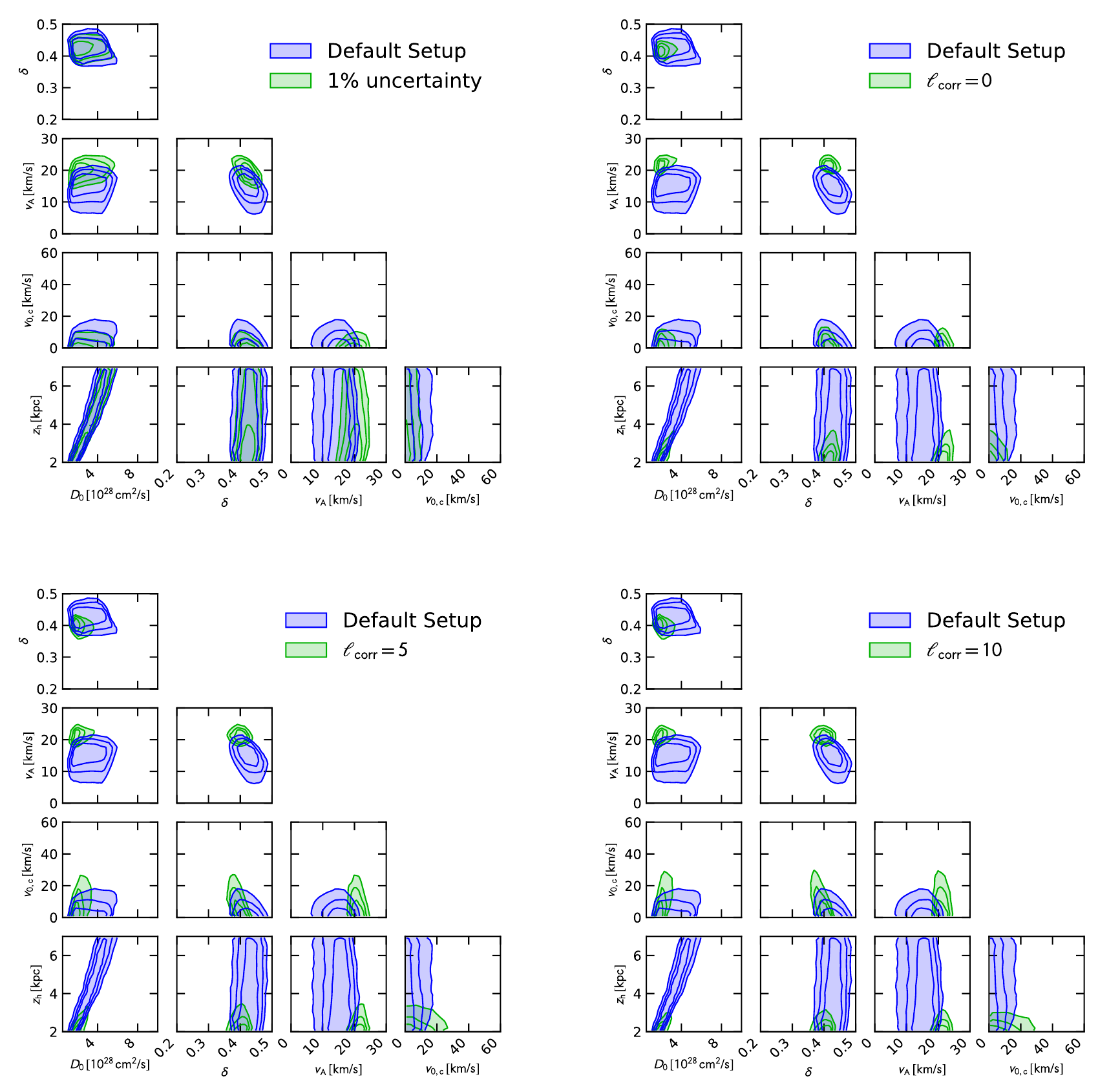}}     
    \end{picture}
	\caption{ Uncertainty contours (1-3$\sigma$) of the propagation parameters corresponding to 
	          the four different treatments of systematic uncertainties and its correlation discussed in 
	          Sec.~\ref{sec::corr}.
	 }
  \label{fig::triangles_corr}
\end{figure}
%                                      \         |
%                                        \       |
%                                          \     |
%=====================

Figure~\ref{fig::corr_datadriven} shows the results of this procedure applied to proton, helium, and antiproton-over-proton data.
The figure shows the likelihood profiles and the corresponding 1$\sigma$ to 3$\sigma$ contours in the frequentist interpretation. 
As can be seen, the resulting constraints on the three parameters are not very strong. 
In particular, the correlation length ${\ell_\text{corr}}$ is not constrained at the 2$\sigma$ level for all three datasets, 
whereas $\alpha$ is constrained to small 
values for  ${\ell_\text{corr}} \gsim 5$ and unconstrained for smaller correlation lengths. 
The only well-constrained quantity is the fraction of the systematic uncertainty with a small correlation length, $f$. 
The fits converge to $f\sim0.07$ and 0.1 for proton and helium, respectively, and $f\lsim 0.3$ for antiprotons .
We note that, interestingly, the potential correlation of all three datasets can be described by the same parameters within uncertainties.
Experimentally, however, it is unclear if equal systematics properties are expected for the three cases.    
Therefore, we will consider the potential correlation independently for each CR species in the following.

Given the large uncertainty in the determination of  ${\ell_\text{corr}}$, and $\alpha$,
we choose three benchmark cases compatible with the constraints to study the effects on the fit results in more detail. 
Specifically, the benchmark scenarios correspond to three different fixed values for the correlation length, ${\ell_\text{corr}}= 0,\, 5$ and 10.
Let us first consider proton and helium. Figure~\ref{fig::corr_datadriven} shows that the values for $f$ and $\alpha$ are constrained at every fixed value of the correlation length. Hence, we fix their values to the maximum of the likelihood profiles at each ${\ell_\text{corr}}$. The exact values are summarized in Tab.~\ref{tab::corr_benchmarks}\@. For the antiproton-to-proton ratio the situation is different for two reasons: First, the number of data points is smaller compared to the proton and helium datasets. Secondly, and probably even more importantly, the size of the statistical uncertainty is significantly larger. While proton and helium are clearly dominated by only systematic uncertainties, the size of systematic and statistical uncertainty is comparable for the $\pb/p$ ratio. As a result, the best-fit $f$ converges towards 0 and is only constrained from above, while $\alpha$ is completely unconstrained. 
Since our goal is to study the effect of possible correlations in the AMS-02 data, choosing $f=0$, i.e., the maximum of the
likelihood is not a  good choice. Instead, to be conservative, we choose the maximum $f$ value allowed
by our constraints. In particular, we choose $f=0.3$, which is approximately the maximum allowed value at a confidence
level of 90\%. For $\alpha$ we choose a representative value of 1. Again, these values are summarized in Tab.~\ref{tab::corr_benchmarks}.

\subsection{Results}

In total, we perform eight fits to investigate the effect of correlations in the AMS-02 data. They correspond to four different setups of the uncertainties where each setup is fitted once with and without DM\@. Summarizing, the four setups are the fixed 1\% systematic uncertainty without any correlation of data points, and the three different benchmark scenarios from the data-driven covariance matrix with correlation lengths 0, 5, and 10.
The results of the various fits are shown in Tab.~\ref{tab::bestfit_parameter3}.
As for the case in which we studied the cross sections, we show the residuals of the antiproton-to-proton ratio in Fig.~\ref{fig::pbar_residuals_corr}. The shown uncertainties are the square roots of the diagonal elements of the covariance matrix and, therefore, some care is needed when drawing conclusions directly from the figures in the cases of ${\ell_\text{corr}}=5$ or 10. In general, however,  we can clearly observe that the uncertainties are considerably smaller with respect to  the default fit. 
From a methodological point of view the smaller uncertainties result in a more complicated fit. In order to converge, the new fits require up to 1.8 million \textsc{MultiNest} likelihood evaluations using the same configuration as in the default setup.
The fit with fixed systematic uncertainty at 1\% converges to a $\chi^2$/dof of $77.4/143\approx0.5$. In more detail, the $\chi^2$ of the AMS-02 antiproton-to-proton ratio is 44 which now matches  well with the 42 $\pb/p$ data points included in the fit, and it is thus reasonable also in terms of  goodness of fit.
 On the other hand, the $\chi^2$'s of proton and helium are 8 and 10, respectively, and are still much smaller than the number of data points, i.e., 50 each.
 This hints to the fact that the uncertainty of antiprotons is estimated reasonably well while the one for proton and helium is still overestimated. 
 The effect on the CR parameter estimation is shown in Fig.~\ref{fig::triangles_corr} (upper left panel): Compared to our default fit, the size of the uncertainty contours slightly shrinks. The only further notable effect is that the Alfv\`en velocity is shifted to slightly larger values, but by less than 1$\sigma$.
 Adding DM in this scenario  improves the fit by a $\Delta\chi^2$ of 30 which would correspond to an evidence of 5.1$\sigma$.

The three benchmark scenarios from the data-driven approach converge to a total $\chi^2$s in the range 250--280, 
which is about a factor of 2 larger than the dof of 143. 
The $\chi^2$ for $\pb/p$ is in the range 40--70, which, again, is reasonable in terms of goodness of fit, especially
considering the DM fits, where the  $\chi^2$ range shrinks to 40--50.
Instead, the main contribution to the total $\chi^2$ comes from $p$ and He which are in the range 80--100.
With an $f\sim 0.1$ for $p$ and He the total error is at the level of $\sim$ 0.2\%, so these high values of $\chi^2$ 
are probably not unreasonable since it is unlikely that the accuracy of the model is at such a high level of precision.
Finally, from the triangle plots in Fig.~\ref{fig::triangles_corr} we see that the size of the uncertainty contours shrinks significantly
with respect to the default fit, an outcome which is expected given the smaller error bars.

%=====================
%    \                                           |
%      \                                         |
%        \                                       |
\begin{figure}[t!]
\setlength{\unitlength}{0.1\textwidth}
\begin{picture}(10,3.5)
\put( 2.50, -0.2){\includegraphics[width=0.5\textwidth]{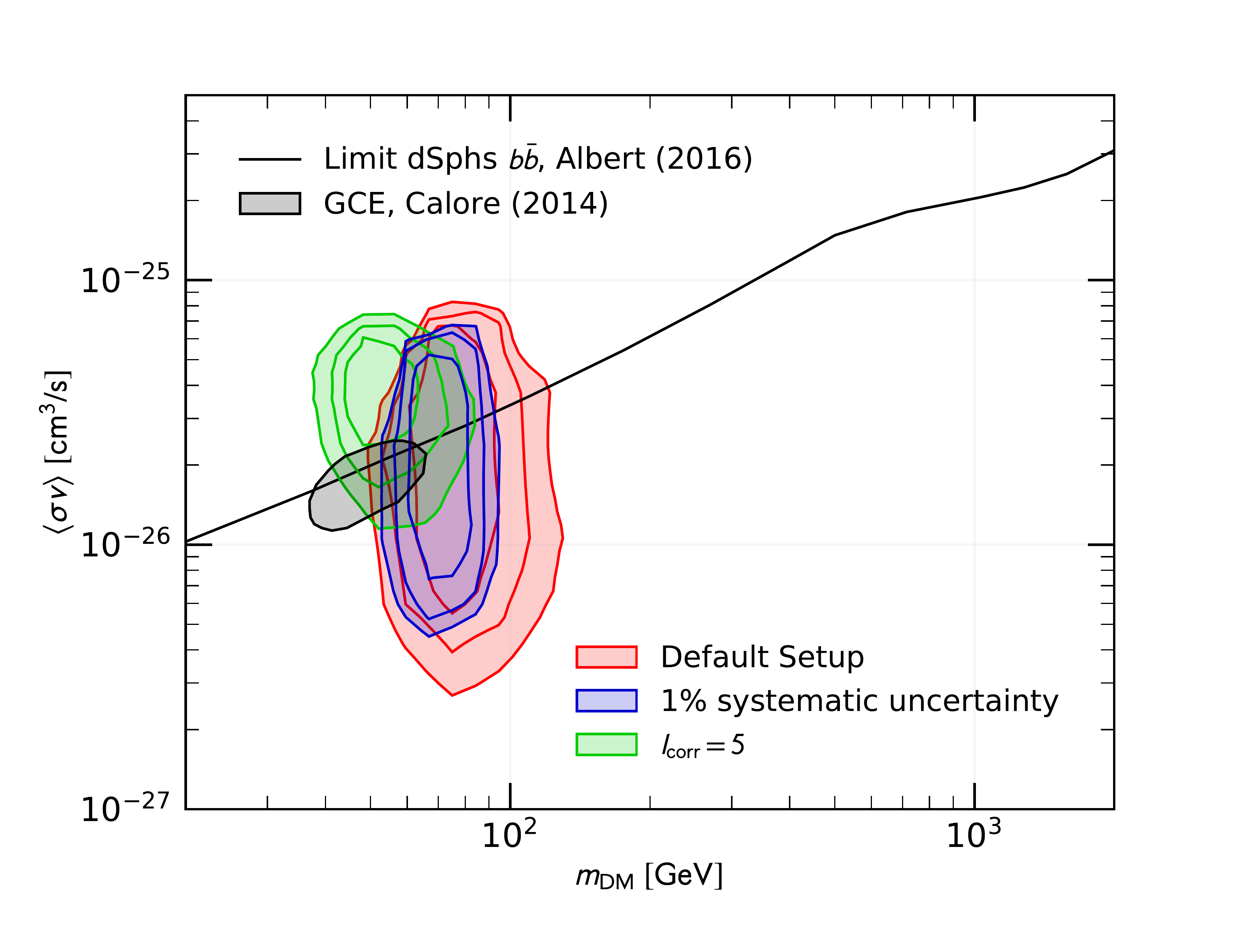}}
\end{picture}
	\caption{
	Contours of the 1--3$\sigma$ best-fit in the plane of DM mass and annihilation cross section.
	We show the result of the default fit and two benchmark scenarios for the correlations:
	1\% fixed systematic uncertainties and ${\ell_\text{corr}}=5$. For comparison we show the 
	limit for the DM annihilation cross section derived from the observation of 
	dwarf spheroidal galaxies~\cite{Fermi-LAT:2016uux} and the $2\sigma$ best-fit region of the 
	GCE~\cite{Calore:2014nla}.
	}
  \label{fig::DM_contours_corr}
\end{figure}
%                                      \         |
%                                        \       |
%                                          \     |
%=====================
 
As already observed in the case of the first method, we notice a shift of $v_\mathrm{A}$ to slightly larger values
although compatible at 1--2$\sigma$ level. Furthermore, there is now a preference for a small halo height which converges towards 2~kpc which is the boundary of the fit range. 
This small value of $z_\mathrm{h}$ is disfavored by observations of electron and positron spectra~\cite{DiMauro:2014iia}, but still possible. The result also shows that with very small errors the fit becomes sensitive to $z_\mathrm{h}$ even without the use of cosmic clock data like ${}^{10}$Be/${}^{9}$Be.
This further stresses the importance of studying in detail the properties of the data to understand if such
small uncertainties can be used and exploited. 
All three benchmarks show evidence for a DM signal. The significance depends on the correlation length, namely, the improvement in the fit quality reaches $\Delta \chi^2=34$ (corresponding to $5.5\sigma$) in the case of ${\ell_\text{corr}}=0$, while it drops to around 18 ($3.8\sigma$) for ${\ell_\text{corr}}=5$ and 10.\footnote{We checked that these significances do not strongly depend on the value for $z_\mathrm{h}$, e.g.~at a fixed value of $z_\mathrm{h}=4$~kpc we get a $\Delta \chi^2$ of 36, 20, and 21 for the cases of ${\ell_\text{corr}}=0$, 5, and 10, respectively.} Again, the results with the above different setups and benchmark scenarios show the importance of modeling correlations correctly. Depending on the ``true'' covariance, results may be driven in different directions, as exemplarily pointed out here for the case of the potential DM signal. In particular, typically, we see that the use of covariance improves the significances and reduces the errors on the estimated parameters, which indicates that the full potential of the data is not yet exploited.

The shrinking of the contours with respect to the case in which the correlation among the data-points is neglected is also observed for the DM mass and velocity-averaged annihilation cross section. Figure~\ref{fig::DM_contours_corr} shows the contours of the benchmark scenarios of 1\% fixed systematic uncertainties and ${\ell_\text{corr}}=5$, respectively. In the latter case, the contour is additionally shifted to slightly smaller values of the DM mass and to larger $\sv$. 
The contours for ${\ell_\text{corr}}=0$ and 10 are very similar to ${\ell_\text{corr}}=5$. 
The fact that the contours do not extent to lower values of $\sv$ as for the default fit case is directly related to the preference of the fit for small $z_\mathrm{h}$ values, which
give a lower DM signal.  Lower values of $\sv$ in the default fit are, correspondingly, associated to larger values of $z_\mathrm{h}$ which provide an increased
normalization of the DM signal.
We remark, nonetheless, that the DM contours of all our fits are compatible with each other. 
We also stress that the contours are derived for a fixed value of the local DM density of $\rho_\odot= 0.43$ GeV/cm$^3$,
as mentioned in Sec.~\ref{sec:cr}, and that the additional uncertainty in $\rho_\odot$ (conservatively in the range 0.2--0.7 GeV/cm$^3$~\cite{Salucci:2010qr}) would contribute to extend to contours to both smaller and larger values of  $\sv$.

Furthermore, we note that a combination of cross-section uncertainties and correlations in the AMS-02 data does not significantly change the picture. We checked this by performing a fit taking into account the uncertainties of the cross-section parametrization (following the joint fit method as introduced in Sec.~\ref{sec:jointxsfit}) as well as taking into account the 1\% uncorrelated uncertainty from AMS-02. As expected, the resulting fit is very similar to the one fixing the cross-section parameters to their best-fit values.

Finally, we remark again that all results of this section have to be taken with caution. By guessing different possible scenarios of the AMS-02 uncertainties and its covariances in data, we demonstrate, merely as a proof of concept, that correlations in data can impose an important effect on the results, in particular, on the significance of the potential DM signal. Nonetheless, we do not deem our four benchmark scenarios as fully representative of the complete set of possible covariances in data or to contain the true covariance matrices. 
Our main aim is, instead, to show the importance of these effects in order to trigger further investigations on the subject, especially from the experimental collaborations, which, through the knowledge of the detector, have handles to study this issue more in detail.
As we have shown, a better characterization of the covariance of the uncertainties can provide much more constraining
power, and thus fully exploit the potential of the experimental data.

%===================================================================
\section{Conclusion} \label{sec::conclusion}
%===================================================================

In this paper we investigated two important sources of uncertainties in the interpretation of 
global CR fits to AMS-02 data: the antiproton production cross section and the correlations in the experimental data.
In particular, we investigated their impact on the potential dark matter signal found in previous studies~\cite{Cuoco:2016eej,Cui:2016ppb,Cuoco:2017rxb}.

First, we reproduced the findings of~\cite{Cuoco:2016eej} in an updated setup using the most recent cross-section parametrization
from~\cite{Korsmeier:2018gcy} and an improved treatment of solar modulation. In addition, we studied a possible tension between the AMS-02 and CREAM data at large rigidities by either excluding CREAM data from the global fit or by taking into account a possible normalization uncertainty. We found that the potential DM signal persists, although with a slightly smaller significance of around $3\sigma$ (with respect to $4.5\sigma$ found in~\cite{Cuoco:2016eej}). The reason for the reduction in significance is mainly due to the updated antiproton production cross-section parametrization.

We studied the uncertainties induced by the antiproton cross-section parametrizations using two approaches. On the one hand, we described the uncertainties by a covariance matrix assuming a multivariate Gaussian distribution. On the other hand, we performed a joint fit of CR and the most relevant cross-section parameters. We found that both descriptions lead to comparable results. As the latter approach reveals, this is due to the absence of strong correlations between the CR and cross-section parameters. We hence concluded that neither the results for the CR propagation parameters nor the significance of the DM signal are strongly affected by the antiproton cross-section uncertainties. 

In all fits the $\chi^2/\text{dof}$ is significantly below one. We expect this to be a result of neglecting correlations in the AMS-02 data, which are not provided by the experimental collaboration. A long-distance correlation (in rigidity bins) amounts to an overall normalization or tilt of the spectrum, and it can be absorbed into the fit parameters. Using the systematic errors as given by AMS-02 and assuming these to be uncorrelated can, hence, greatly overestimate the error. We followed different approaches in order to illustrate the importance of the knowledge of these correlations. On the one hand, we simply reduced the uncorrelated error to 1\% assuming the remaining error to be fully correlated (and absorbed in the overall normalization). As a result the overall fit quality worsens slightly, however, still providing $\chi^2/\text{dof}$ well below one. Interestingly, the preference for DM becomes more significant reaching $5\sigma$. 
On the other hand, we followed a data-driven approach, determining the maximal fraction of a correlated uncertainty for three choices of the correlation length (in terms of the distance between rigidity bins), $\ell_\text{corr}= 0,\,5,\,10$.
Our analysis shows that only a small fraction of the systematic uncertainty can be correlated this way,
namely, about 10\% for the $p$ and He data and no more than 30\% (at 90\% C.L.) for the $\pb/p$ data. The 
remainder of the systematic uncertainty is compatible with just an overall shift of the global normalization.
However, the identification of the exact shape of a covariance matrix from data and a typical correlation length is not unique. 
As an illustration and as a proof of concept, we choose three benchmark scenarios, although these are not expected to cover the whole
ranges of possibilities. A more realistic characterization of the uncertainties can only be achieved in the future with further insight from the AMS-02 detector and analyses. Since the fraction of short-range correlated uncertainties is very small, for the moment treating AMS-02 systematic uncertainties as uncorrelated (i.e.\ statistical) is conservative.  Nonetheless, we have shown that with a proper treatment of the uncertainties, the data can be much more constraining,
both on the determination of propagation parameters and on the presence of a signal for DM\@.
In particular, we find that the significance of a  DM annihilation signals increases up to 5.5$\sigma$
depending on how we model the correlation.
We thus stress that getting this kind of insight in the future would be extremely valuable and would allow one to better exploit the potential of the experimental data.  

Finally, we note that two aspects remain to be studied in more detail in the future. First, solar modulation is not fully understood for rigidities below about $5$~GV and has to be investigated using the recently published monthly data of AMS-02. In addition, different diffusion models have to be studied to allow to include other secondary CR data such as ${}^3$He, Be, Li and B into our global analysis.

%===================================================================
\section*{\label{sec::acknowledgments}Acknowledgments}
%===================================================================

We thank Henning Gast, Bj{\"o}rn Sarrazin and Stefan Schael for helpful discussions and comments. 
J.H. acknowledges support from the F.R.S.-FNRS, of which he is a postdoctoral researcher.
The work of M. Korsmeier is supported by the ``Departments of Excellence 2018-2022'' grant awarded by
the Italian Ministry of Education, University and Research (MIUR) (L.~232/2016).
Simulations were performed with computing resources granted
by RWTH Aachen University under projects rwth0085 and rwth0272.
 
 \medskip 
\emph{Note added:}
Shortly after this analysis was made public, Refs.~\cite{Cholis:2019ejx,Lin:2019ljc} appeared on the arXiv discussing similar subjects.
The authors of \cite{Cholis:2019ejx} investigate the impact of cross-section uncertainties as well as time, charge and energy-dependent solar modulation effects on the significance of the DM signal, and arrive at conclusions which are similar to ours. In \cite{Lin:2019ljc} the authors exploit machine learning to marginalize more efficiently over cosmic-ray propagation uncertainties.

\bibliography{bibliography}{}
\bibliographystyle{apsrev4-1.bst}

%=====================================================================
%    \                                                                                        |
%      \                                                                                      |
%        \                                                                                    |
\begin{table}
  \scriptsize
  \caption{  The best-fit parameters of the fits with and without DM\@. The double-column contains the results of
             our default setup. We show for comparison a fit that includes CREAM data and AMS-02 above 300 GeV.}
  \centering
  \begin{tabular}{ c c c @{\hspace{5px}} c c } \hline\hline
    \textbf{parameter}                                        & \multicolumn{2}{c }{\textbf{default setup}} & \multicolumn{2}{c }{\textbf{extended setup}}                                                                                                                                                                                                                    \\ \hline
    XS parametrization                                        & \multicolumn{2}{c}{param.~MW}                         &  \multicolumn{2}{c}{param.~MW}                                                 \\
    DM                                                        & incl.                     & excl.                     &  incl.                               &    excl.                                \\
    $\gamma_{1}$                                              & $1.71^{+0.02}_{-0.25}$    & $1.72^{+0.04}_{-0.12}$    &   $ 1.64^{+0.05}_{-0.12} $           &    $ 1.72^{+0.02}_{-0.11}       $       \\
    $\gamma_{1,p}$                                            & $1.78^{+0.003}_{-0.19}$   & $1.75^{+0.03}_{-0.10}$    &   $ 1.73^{+0.04}_{-0.06} $           &    $ 1.73^{+0.05}_{-0.07}       $       \\
    $\gamma_{2}$                                              & $2.41^{+0.03}_{-0.002}$   & $2.38^{+0.01}_{-0.02}$    &   $ 2.44^{+0.01}_{-0.02} $           &    $ 2.38^{+0.01}_{-0.01}       $       \\
    $\gamma_{2,p}$                                            & $2.45^{+0.03}_{-0.002}$   & $2.42^{+0.01}_{-0.02}$    &   $ 2.48^{+0.01}_{-0.02} $           &    $ 2.41^{+0.01}_{-0.01}       $       \\
    $R_{0},\ \mathrm{[MV]}$                                   & $6950^{+330}_{-1640}$     & $7380^{+910}_{-1450}$     &   $ 6519^{+1045}_{-824}  $           &    $ 7695^{+563}_{-1375}        $       \\
    $s_{0}$                                                   & $0.38^{+0.06}_{-0.04}$    & $0.34^{+0.05}_{-0.04}$    &   $ 0.38^{+0.06}_{-0.01} $           &    $ 0.37^{+0.04}_{-0.03}       $       \\
    $D_{0},\ \mathrm{[10^{28}\,cm^{2}/s]}$                    & $5.43^{+0.45}_{-3.17}$    & $2.90^{+1.33}_{-1.21}$    &   $ 3.46^{+2.46}_{-1.19} $           &    $ 1.97^{+1.27}_{-3.81}       $       \\
    $\delta$                                                  & $0.38^{+0.01}_{-0.03}$    & $0.42^{+0.02}_{-0.01}$    &   $ 0.35^{+0.02}_{-0.01}  $          &    $ 0.42^{+0.01}_{-0.02}       $       \\
    $v_{\rm A},\ \mathrm{[km/h]}$                             & $18.0^{+2.1}_{-1.4}$      & $16.2^{+1.0}_{-2.5}$      &   $ 18.59^{+0.00}_{-3.25} $          &    $ 15.81^{+0.87}_{-1.99}      $       \\
    $v_{0,{\rm c}},\ \mathrm{[km/h]}$                         & $0.08^{+9.09}_{-0.08}$    & $0.52^{+2.32}_{-0.51}$    &   $ 0.35^{+4.94}_{-0.14}  $          &    $ 0.79^{+2.19}_{-0.77}       $       \\
    $z_{\rm h},\ \mathrm{[kpc]}$                              & $6.45^{+0.30}_{-4.26}$    & $3.58^{+2.36}_{-1.52}$    &   $ 3.36^{+3.47}_{-1.13}  $          &    $ 2.47^{+1.58}_{-0.43}       $       \\
    $\log(m_{\rm DM}/\mathrm{[GeV]})$                         & $1.89^{+0.03}_{-0.08}$    &                           &   $ 1.88^{+0.05}_{-0.00}   $         &                                         \\
    $\log(\langle\sigma v\rangle/\mathrm{[s/cm^{3}]})$        & $-26.16^{+0.78}_{-0.04}$  &                           &   $ -25.56^{+0.20}_{-0.47}$          &                                         \\
    $\delta_2$                                                &                           &                           &   $ 0.23^{+0.01}_{-0.00} $           &    $ 0.30^{+0.02}_{-0.02}       $       \\
    $R_{1},\ \mathrm{[GV]}$                                   &                           &                           &   $ 344^{+26}_{-20} $                &    $ 338^{+49}_{-40}            $       \\
    $\varphi_{\text{SM,AMS-02},p,{\rm He}},\ \mathrm{[MV]}$   &  $616^{+71}_{-72}$        & $625^{+55}_{-85}$         &   $ 566^{+19}_{-71} $                &    $ 567^{+19}_{-66}            $       \\
    $\varphi_{\text{SM,AMS-02},\pb},\ \mathrm{[MV]}       $   &  $604^{+112}_{-114}$      & $561^{+135}_{-112}$       &   $ 577^{+43}_{-82} $                &    $ 561^{+35}_{-106}           $       \\
    $\chi_{\text{AMS-02},p  }^{2}$                            &  3.2                      &  2.6                      &   4.9                                &     7.0                                 \\
    $\chi_{\text{AMS-02,He} }^{2}$                            &  4.0                      &  4.8                      &   10.4                               &    12.0                                 \\
    $\chi_{\text{AMS-02},\pb}^{2}$                            & 11.1                      & 22.1                      &   12.2                               &    20.0                                 \\
    $\chi_{\text{Voager},p  }^{2}$                            &  3.2                      &  3.8                      &   2.7                                &     3.6                                 \\
    $\chi_{\text{Voager,He} }^{2}$                            &  1.3                      &  1.9                      &   3.4                                &     4.1                                 \\
    $\chi_{\text{CREAM},p  }^{2}$                             &                           &                           &   1.0                                &     1.0                                 \\
    $\chi_{\text{CREAM,He} }^{2}$                             &                           &                           &   1.0                                &     2.9                                 \\
    $\chi_{\varphi_\text{SM}   }^{2}$                         &  0.0                      &  0.4                      &   0.0                                &     0.0                                 \\
    $\chi_{\text{CR}        }^{2}$                            & 22.9                      & 35.6                      &   35.6                               &    50.7                                 \\
    $\chi^{2}$/dof                                            & 22.9/143                  & 35.6/145                  &   35.6/177                           &    50.7/179                             \\ \hline
  $\Delta \chi^2$                                             & \multicolumn{2}{c}{12.7}                              &  \multicolumn{2}{c}{15.1}                                                      \\
  DM significance                                             & \multicolumn{2}{c}{$3.1\sigma$}                       &  \multicolumn{2}{c}{$3.5\sigma$}                                               \\
    \hline\hline
  \end{tabular} 
    \label{tab::bestfit_parameter1}
\end{table}
%                                                                                  \         |
%                                                                                    \       |
%                                                                                      \     |
%=====================================================================

%=====================================================================
%    \                                                                                        |
%      \                                                                                      |
%        \                                                                                    |
\begin{table}
  \scriptsize
  \caption{The best-fit parameters of various fits to test the impact of cross-section uncertainties. For details refer to Section~\ref{sec::xs}. }
  \centering
  \begin{tabular}{ c @{\hspace{5px}} c c @{\hspace{10px}} c c @{\hspace{5px}} c c @{\hspace{10px}} c c} \hline\hline
    \textbf{parameter}                                        & \multicolumn{2}{c }{\textbf{default setup, different XS}}    & \multicolumn{4}{c }{\textbf{XS uncertainty by covariance matrix}}                                     & \multicolumn{2}{c}{\textbf{joint fit (CR$+$XS)}}      \\ \hline
    XS parametrization                                        & \multicolumn{2}{c}{param.~MD}                                & \multicolumn{2}{c}{param.~MW}                      & \multicolumn{2}{c}{param.~MD}                    & \multicolumn{2}{c}{param.~MW}                         \\
    DM                                                        & incl.                      & excl.                           & incl.                   & excl.                    & incl.                   & excl.                  & incl.                    & excl.                      \\
    $\gamma_{1}$                                              & $1.55^{+0.11}_{-0.05}$     & $1.65^{+0.09}_{-0.09}$          & $1.65^{+0.07}_{-0.21}$  & $1.68^{+0.07}_{-0.09}$   & $1.52^{+0.08}_{-0.06}$  & $1.64^{+0.09}_{-0.06}$ & $1.57^{+0.06}_{-0.05}$   & $1.67^{+0.09}_{-0.09}$     \\
    $\gamma_{1,p}$                                            & $1.67^{+0.08}_{-0.02}$     & $1.69^{+0.09}_{-0.07}$          & $1.76^{+0.03}_{-0.17}$  & $1.73^{+0.05}_{-0.08}$   & $1.64^{+0.08}_{-0.05}$  & $1.69^{+0.07}_{-0.05}$ & $1.68^{+0.06}_{-0.05}$   & $1.70^{+0.07}_{-0.05}$     \\
    $\gamma_{2}$                                              & $2.43^{+0.01}_{-0.01}$     & $2.39^{+0.01}_{-0.02}$          & $2.43^{+0.02}_{-0.02}$  & $2.37^{+0.02}_{-0.01}$   & $2.43^{+0.02}_{-0.01}$  & $2.28^{+0.01}_{-0.01}$ & $2.44^{+0.01}_{-0.03}$   & $2.38^{+0.01}_{-0.03}$     \\
    $\gamma_{2,p}$                                            & $2.48^{+0.01}_{-0.01}$     & $2.42^{+0.01}_{-0.02}$          & $2.47^{+0.02}_{-0.02}$  & $2.40^{+0.02}_{-0.01}$   & $2.47^{+0.02}_{-0.01}$  & $2.41^{+0.01}_{-0.01}$ & $2.48^{+0.01}_{-0.03}$   & $2.42^{+0.01}_{-0.03}$     \\
    $R_{0},\ \mathrm{[MV]}$                                   & $5860^{+870}_{-300}$       & $6700^{+1490}_{-620}$           & $6910^{+470}_{-1710}$   & $6880^{+1200}_{-940}$    & $5840^{+510}_{-430}$    & $6830^{+730}_{-130}$   & $5950^{+640}_{-370}$     & $6910^{+970}_{-890}$       \\
    $s_{0}$                                                   & $0.43^{+0.01}_{-0.08}$     & $0.41^{+0.02}_{-0.06}$          & $0.38^{+0.10}_{-0.02}$  & $0.36^{+0.04}_{-0.06}$   & $0.41^{+0.04}_{-0.03}$  & $0.38^{+0.02}_{-0.04}$ & $0.43^{+0.01}_{-0.08}$   & $0.37^{+0.04}_{-0.05}$     \\
    $D_{0},\ \mathrm{[10^{28}\,cm^{2}/s]}$                    & $2.70^{+2.72}_{-0.25}$     & $2.10^{+0.89}_{-0.50}$          & $5.64^{+0.50}_{-3.60}$  & $2.94^{+0.68}_{-1.06}$   &  $2.27^{+0.90}_{-0.13}$ & $1.86^{+0.61}_{-0.19}$ & $2.47^{+4.21}_{-0.01}$   & $1.72^{+1.81}_{-0.26}$     \\
    $\delta$                                                  & $0.35^{+0.02}_{-0.01}$     & $0.41^{+0.02}_{-0.02}$          & $0.35^{+0.03}_{-0.02}$  & $0.43^{+0.01}_{-0.03}$   & $0.35^{+0.02}_{-0.02}$  & $0.42^{+0.02}_{-0.01}$ & $0.35^{+0.04}_{-0.02}$   & $0.42^{+0.02}_{-0.0004}$   \\
    $v_{\rm A},\ \mathrm{[km/h]}$                             & $16.8^{+2.7}_{-1.8}$       & $14.6^{+1.9}_{-1.7}$            & $19.5^{+0.5}_{-4.9}$    & $15.0^{+2.1}_{-0.6}$     & $17.2^{+2.7}_{-2.2}$    & $14.3^{+1.4}_{-1.8}$   & $17.9^{+0.3}_{-2.2}$ &   $15.2^{+0.66}_{-1.38}$       \\
    $v_{0,{\rm c}},\ \mathrm{[km/h]}$                         & $4.80^{+2.44}_{-4.41}$     & $1.97^{+4.11}_{-1.95}$          & $2.63^{+6.21}_{-2.44}$  & $0.66^{+1.65}_{-0.61}$   & $4.22^{+3.67}_{-3.66}$  & $2.78^{+1.01}_{-2.40}$ & $1.17^{+5.39}_{-0.15}$   & $0.27^{+2.02}_{-0.19}$     \\
    $z_{\rm h},\ \mathrm{[kpc]}$                              & $2.62^{+2.49}_{-0.23}$     & $2.54^{+1.24}_{-0.53}$          & $5.76^{+0.50}_{-3.70}$  & $3.72^{+0.82}_{-1.48}$   & $2.20^{+0.81}_{-0.14}$  & $2.28^{+0.86}_{-0.22}$ & $2.46^{+4.21}_{-0.11}$   & $2.15^{+2.61}_{-0.15}$     \\
    $\log(m_{\rm DM}/\mathrm{[GeV]})$                         & $1.85^{+0.02}_{-0.05}$     &                                 & $1.88^{+0.07}_{-0.06}$  &                          & $1.82^{+0.04}_{-0.03}$  &                        & $1.89^{+0.04}_{-0.03}$   &                            \\
    $\log(\langle\sigma v\rangle/\mathrm{[s/cm^{3}]})$        & $-25.50^{+0.01}_{-0.36}$   &                                 & $-25.88^{+0.59}_{-0.09}$&                          & $-25.32^{+0.01}_{-0.41}$&                        & $-25.47^{+0.10}_{-0.42}$ &                            \\
    $\varphi_{\text{SM,AMS-02},p,{\rm He}},\ \mathrm{[MV]}$   & $554^{+101}_{-13}$         &  $544^{+44}_{-58}$              & $582^{+90}_{-64}$       &  $582^{+54}_{-25}$       &  $582^{+38}_{-46}$      &  $556^{+7}_{-28}$      & $584^{+39}_{-80}$        &  $598^{+22}_{-61}$         \\
    $\varphi_{\text{SM,AMS-02},\pb},\ \mathrm{[MV]}       $   & $549^{+31}_{-45}$          &  $522^{+62}_{-115}$             & $592^{+120}_{-90}$      &  $605^{+56}_{-114}$      &  $600^{+54}_{-56}$      &  $553^{+17}_{-55}$     & $498^{+166}_{-81}$       & $537^{+152}_{-61}$         \\
    $C_{1},\ \mathrm{[10^{-3}\ (GeV)^{-2}]}$                  &                            &                                 &                         &                          &                         &                        & $50.1^{+0.5}_{-1.0} $    & $51.0^{+0.5}_{-1.7}$       \\
    $C_{5},\ \mathrm{[10^{-3}]}$                              &                            &                                 &                         &                          &                         &                        & $0.32^{+0.24}_{-0.02}$   & $0.27^{+0.08}_{-0.17}$     \\
    $C_{6}$                                                   &                            &                                 &                         &                          &                         &                        & $3.82^{+0.27}_{-0.40}$   &$3.73^{+0.81}_{-0.10}$      \\
    $\chi_{\text{AMS-02},p  }^{2}$                            & 2.4                        & 3.8                             & 2.6                     &  2.6                     & 2.2                     & 2.6                    & 2.8                      & 2.6                        \\
    $\chi_{\text{AMS-02,He} }^{2}$                            & 4.4                        & 4.6                             & 4.1                     &  5.1                     & 4.1                     & 4.3                    & 3.9                      & 4.7                        \\
    $\chi_{\text{AMS-02},\pb}^{2}$                            & 11.0                       & 21.1                            & 11.4                    & 21.3                     & 11.3                    & 22.4                   & 11.7                     & 19.9                       \\
    $\chi_{\text{Voager},p  }^{2}$                            & 3.5                        & 3.7                             & 3.2                     &  3.7                     & 3.1                     & 3.8                    & 2.8                      & 3.5                        \\
    $\chi_{\text{Voager,He} }^{2}$                            & 0.9                        & 1.0                             & 1.7                     &  1.1                     & 1.1                     & 1.0                    & 1.1                      & 1.9                        \\
    $\chi_{\varphi_\text{SM}   }^{2}$                         & 0.0                        & 0.1                             & 0.0                     &  0.1                     & 0.0                     & 0.0                    & 0.7                      & 0.4                        \\
    $\chi_{\text{CR}        }^{2}$                            & 22.2                       & 34.2                            & 23.0                    & 33.9                     & 22.0                    & 34.2                   & 23.2                     & 33.0                       \\
    $\chi_{\text{XS}        }^{2}$                            &                            &                                 &                         &                          &                         &                        & 791.4                    & 792.2                      \\
    $\chi^{2}$/dof                                            & 22.2/143                   & 34.2/145                        & 23.0/143                & 33.9/145                 & 22.0/143                & 34.2/145               & 814.5/799                & 825.2/801                  \\ \hline
    $\Delta \chi^2$                                           & \multicolumn{2}{c}{12.0}                                     & \multicolumn{2}{c}{10.9}                           & \multicolumn{2}{c}{12.2}                         &  \multicolumn{2}{c}{10.7}                             \\
    DM significance                                           & \multicolumn{2}{c}{$3.0\sigma$}                              & \multicolumn{2}{c}{$2.9\sigma$}                    & \multicolumn{2}{c}{$3.1\sigma$}                  &  \multicolumn{2}{c}{$2.8\sigma$}                      \\
   \hline\hline
  \end{tabular}
  \label{tab::bestfit_parameter2}
\end{table}

%                                                                                  \         |
%                                                                                    \       |
%                                                                                      \     |
%=====================================================================

%=====================================================================
%    \                                                                                        |
%      \                                                                                      |
%        \                                                                                    |
\begin{table}
  \scriptsize
  \caption{The best-fit parameters of various fits to test the impact of correlated AMS-02 uncertainties. For details refer to Section~\ref{sec::corr}.}
  \centering
  \begin{tabular}{ c c c @{\hspace{20px}} c c @{\hspace{10px}} c c @{\hspace{10px}} c c @{\hspace{10px}} c c } \hline\hline
     \textbf{parameter}                                     & \multicolumn{2}{ c }{\textbf{1\% $\bm{\sigma_\text{sys}$}}}    & \multicolumn{2}{c}{$\bm{\ell_\text{corr}=0}$}                          & \multicolumn{2}{c}{$\bm{\ell_\text{corr}=5}$}                 & \multicolumn{2}{c}{$\bm{\ell_\text{corr}=10}$}                  \\ \hline
     XS parametrization                                     & \multicolumn{2}{c}{param.~MW}                                  & \multicolumn{2}{c}{param.~MW}                                          & \multicolumn{2}{c}{param.~MW}                                 & \multicolumn{2}{c}{param.~MW}                                   \\
     DM                                                     & incl.                       &  excl.                           & incl.                              & excl.                             & incl.                        & excl.                          & incl.                          & excl.                          \\
     $\gamma_{1}$                                           & $1.62^{+0.08}_{-0.10}$      &  $1.77^{+0.02}_{-0.08}$          &   $ 1.65^{+0.02}_{-0.03} $         & $ 1.70^{+0.03}_{-0.05} $          &  $ 1.61^{+0.10}_{-0.12} $    & $ 1.72^{+0.02}_{-0.06} $       &  $ 1.65^{+0.07}_{-0.11} $      & $ 1.67^{+0.03}_{-0.06} $       \\
     $\gamma_{1,p}$                                         & $1.70^{+0.08}_{-0.07}$      &  $1.74^{+0.02}_{-0.07}$          &   $ 1.67^{+0.01}_{-0.04} $         & $ 1.66^{+0.02}_{-0.04} $          &  $ 1.64^{+0.09}_{-0.13} $    & $ 1.69^{+0.02}_{-0.06} $       &  $ 1.66^{+0.04}_{-0.13} $      & $ 1.65^{+0.02}_{-0.06} $       \\
     $\gamma_{2}$                                           & $2.43^{+0.01}_{-0.01}$      &  $2.37^{+0.02}_{-0.002}$         &   $ 2.42^{+0.00}_{-0.01} $         & $ 2.39^{+0.01}_{-0.00} $          &  $ 2.43^{+0.00}_{-0.01} $    & $ 2.40^{+0.00}_{-0.00} $       &  $ 2.42^{+0.01}_{-0.01} $      & $ 2.39^{+0.01}_{-0.00} $       \\
     $\gamma_{2,p}$                                         & $2.47^{+0.01}_{-0.01}$      &  $2.40^{+0.02}_{-0.001}$         &   $ 2.45^{+0.00}_{-0.01} $         & $ 2.41^{+0.01}_{-0.00} $          &  $ 2.46^{+0.00}_{-0.02} $    & $ 2.42^{+0.00}_{-0.00} $       &  $ 2.45^{+0.01}_{-0.01} $      & $ 2.42^{+0.01}_{-0.00} $       \\
     $R_{0},\ \mathrm{[MV]}$                                & $6600^{+720}_{-1090}$       &  $9180^{+50}_{-1270}$            &   $ 6869^{+200}_{-173}   $         & $ 7844^{+575}_{-512}   $          &  $ 6519^{+1300}_{-793 } $    & $ 7983^{+340}_{-707}   $       &  $ 6743^{+1103}_{-743 } $      & $ 7290^{+262 }_{-582 } $       \\
     $s_{0}$                                                & $0.42^{+0.03}_{-0.05}$      &  $0.32^{+0.06}_{-0.01}$          &   $ 0.45^{+0.01}_{-0.01} $         & $ 0.40^{+0.02}_{-0.01} $          &  $ 0.47^{+0.01}_{-0.04} $    & $ 0.41^{+0.02}_{-0.01} $       &  $ 0.46^{+0.02}_{-0.03} $      & $ 0.41^{+0.02}_{-0.01} $       \\
     $D_{0},\ \mathrm{[10^{28}\,cm^{2}/s]}$                 & $2.80^{+2.35}_{-0.85}$      &  $2.19^{+0.36}_{-0.58}$          &   $ 1.99^{+0.05}_{-0.02} $         & $ 1.79^{+0.34}_{-0.013}$          &  $ 1.94^{+0.22}_{-0.20} $    & $ 1.90^{+0.05}_{-0.03} $       &  $ 1.84^{+0.22}_{-0.02} $      & $ 1.66^{+0.25}_{-0.00}  $      \\
     $\delta$                                               & $0.36^{+0.02}_{-0.01}$      &  $0.43^{+0.002}_{-0.02}$         &   $ 0.37^{+0.01}_{-0.01} $         & $ 0.42^{+0.00}_{-0.01} $          &  $ 0.36^{+0.02}_{-0.00} $    & $ 0.40^{+0.00}_{-0.00} $       &  $ 0.37^{+0.01}_{-0.01} $      & $ 0.41^{+0.00}_{-0.01} $       \\
     $v_{\rm A},\ \mathrm{[km/h]}$                          & $21.1^{+0.4}_{-3.4}$        &  $19.6^{+1.2}_{-1.4}$            &   $21.87^{+0.17}_{-0.17} $         & $ 20.93^{+1.50}_{-0.45}$          &  $ 22.23^{+1.89}_{-1.77}$    & $ 22.15^{+0.15}_{-0.93}$       &  $21.37^{+1.51}_{-0.79} $      & $ 20.25^{+1.41}_{-0.17}$       \\
     $v_{0,{\rm c}},\ \mathrm{[km/h]}$                      & $0.52^{+4.62}_{-0.27}$      &  $0.06^{+0.79}_{-1.36}$          &   $ 0.53^{+2.75}_{-0.27} $         & $ 0.37^{+1.51}_{-0.26} $          &  $ 3.69^{+10.95}_{-3.66}$    & $ 0.24^{+2.65}_{-0.21} $       &  $ 2.01^{+9.99}_{-1.92} $      & $ 0.25^{+3.96}_{-0.00}  $      \\
     $z_{\rm h},\ \mathrm{[kpc]}$                           & $2.76^{+2.68}_{-0.75}$      &  $2.72^{+0.44}_{-0.70}$          &   $ 2.03^{+0.08}_{-0.00} $         & $ 2.05^{+0.41}_{-0.02} $          &  $ 2.00^{+0.29}_{-0.00} $    & $ 2.11^{+0.03}_{-0.01} $       &  $ 2.01^{+0.12}_{-0.00} $      & $ 2.02^{+0.13}_{-0.02} $       \\
     $\log(m_{\rm DM}/\mathrm{[GeV]})$                      & $1.82^{+0.05}_{-0.02}$      &                                  &   $ 1.70^{+0.04}_{-0.01} $         &                                   &  $ 1.74^{+0.01}_{-0.06} $    &                                &  $ 1.71^{+0.05}_{-0.06} $      &                                \\
     $\log(\langle\sigma v\rangle/\mathrm{[s/cm^{3}]})$     & $-25.57^{+0.26}_{-0.37}$    &                                  &   $-25.49^{+0.07}_{-0.12}$         &                                   &  $-25.41^{+0.10}_{-0.10}$    &                                &  $-25.38^{+0.05}_{-0.14}$      &                                \\
     $\varphi_{\text{SM,AMS-02},p,{\rm He}},\ \mathrm{[MV]}$&  $674^{+9}_{-138}$          &   $738^{+58}_{-39}$              &   $819^{+14}_{-0}$                 &  $ 857^{+16}_{-25} $              &  $839^{+20}_{-43}$           &   $844^{+14}_{-26}$            &  $837^{+18}_{-12}$             &  $832^{+26}_{- 2}$             \\
     $\varphi_{\text{SM,AMS-02},\pb},\ \mathrm{[MV]}       $&  $601^{+146}_{-138}$        &   $613^{+69}_{-94}$              &   $741^{+71}_{-81}$                &  $ 568^{+30}_{-55}$               &  $743^{+190}_{-43}$          &   $534^{+57}_{-26}$            &  $791^{+92}_{-99}$             &  $556^{+42}_{-34}$             \\
     $\chi_{\text{AMS-02},p  }^{2}$                         &  3.3                        &   7.6                            &     83.5                           &  81.0                             &  74.4                        &  75.9                          &  79.4                          &  84.3                          \\
     $\chi_{\text{AMS-02,He} }^{2}$                         &  7.6                        &   9.8                            &     97.4                           &  98.9                             &  96.5                        &  94.5                          &  96.7                          &  89.7                          \\
     $\chi_{\text{AMS-02},\pb}^{2}$                         &  30.3                       &  44.0                            &     52.2                           &  69.2                             &  42.3                        &  50.1                          &  47.4                          &  53.9                          \\
     $\chi_{\text{Voager},p  }^{2}$                         &  3.4                        &   7.5                            &      8.8                           &  14.7                             &   7.7                        &   9.5                          &   6.8                          &  13.4                          \\
     $\chi_{\text{Voager,He} }^{2}$                         &  2.1                        &   6.8                            &      7.3                           &  11.4                             &   9.3                        &  10.1                          &  10.4                          &  10.0                          \\
     $\chi_{\varphi_\text{SM}   }^{2}$                      &  0.6                        &   1.7                            &      0.9                           &  8.8                              &   2.4                        &  10.0                          &   0.6                          &   8.0                          \\
     $\chi^{2}$/dof                                         &  47.4/145                   & 77.4/143                         & 250.0/143                          & 284.1/145                         & 232.6/143                    & 250.2/145                      & 241.3/143                      & 259.3/145                      \\
     \hline
       $\Delta \chi^2$                                      & \multicolumn{2}{c}{30.0}                                       & \multicolumn{2}{c}{34.1}                                               &  \multicolumn{2}{c}{17.6}                                     &  \multicolumn{2}{c}{18.0}            \\
  DM significance                                           & \multicolumn{2}{c}{$5.1\sigma$}                                & \multicolumn{2}{c}{$5.5\sigma$}                                        &  \multicolumn{2}{c}{$3.8\sigma$}                              &  \multicolumn{2}{c}{$3.8\sigma$}     \\
   \hline\hline
  \end{tabular}
  \label{tab::bestfit_parameter3}
\end{table}
%                                                                                  \         |
%                                                                                    \       |
%                                                                                      \     |
%=====================================================================

\end{document}